\documentclass{article}
\usepackage{makecell}
\usepackage{float}
\usepackage{amsmath}
\DeclareUnicodeCharacter{2606}{}
\usepackage{multirow}
\usepackage{arxiv}

\usepackage[utf8]{inputenc} 
\usepackage[T1]{fontenc}    
\usepackage{hyperref}       
\usepackage{url}            
\usepackage{booktabs}       
\usepackage{amsfonts}       
\usepackage{nicefrac}       
\usepackage{microtype}      
\usepackage{lipsum}		
\usepackage{graphicx}
\usepackage{natbib}
\usepackage{doi}

\title{Research Data in Scientific Publications: A Cross-Field Analysis}

\author{ Puyu Yang \\
	Institute for Logic, Language and Computation (ILLC)\\
	University of Amsterdam\\
    1098XH, Amsterdam, The Netherlands.\\
	\texttt{p.yang2@uva.nl} \\
	\And
	Giovanni Colavizza\thanks{Giovanni Colavizza is also affiliated at the University of Bologna, Department of Classical and Italian Philology, Italy} \\
	Department of Communication\\
	University of Copenhagen\\
	Karen Blixens Plads 8, Copenhagen, Danmark \\
	\texttt{colavizza@hum.ku.dk} \\
}

\renewcommand{\headeright}{}
\renewcommand{\shorttitle}{Research Data in Scientific Publications: A Cross-Field Analysis}

\hypersetup{
pdftitle={A template for the arxiv style},
pdfsubject={q-bio.NC, q-bio.QM},
pdfauthor={David S.~Hippocampus, Elias D.~Striatum},
pdfkeywords={First keyword, Second keyword, More},
}

\begin{document}
\maketitle

\begin{abstract}
	Data sharing is fundamental to scientific progress, enhancing transparency, reproducibility, and innovation across disciplines. Despite its growing significance, the variability of data-sharing practices across research fields remains insufficiently understood, limiting the development of effective policies and infrastructure. This study investigates the evolving landscape of data-sharing practices, specifically focusing on the intentions behind data release, reuse, and referencing. Leveraging the PubMed open dataset, we developed a model to identify mentions of datasets in the full-text of publications. Our analysis reveals that data release is the most prevalent sharing mode, particularly in fields such as Commerce, Management, and the Creative Arts. In contrast, STEM fields, especially the Biological and Agricultural Sciences, show significantly higher rates of data reuse. However, the humanities and social sciences are slower to adopt these practices. Notably, dataset referencing remains low across most disciplines, suggesting that datasets are not yet fully recognized as research outputs. A temporal analysis highlights an acceleration in data releases after 2012, yet obstacles such as data discoverability and compatibility for reuse persist. Our findings can inform institutional and policy-level efforts to improve data-sharing practices, enhance dataset accessibility, and promote broader adoption of open science principles across research domains.
\end{abstract}


\section{Introduction}
Open Science, emerging from a diverse array of cultural and technological initiatives at the turn of the twenty-first century, has evolved into a transformative movement within the scientific community~\citep{willinsky2005unacknowledged, moore2017genealogy}. At its core, Open Science aims to make scientific research more transparent, accessible, and inclusive, fostering collaboration across disciplines and encouraging broader societal engagement. One influential definition characterizes Open Science as “transparent and accessible knowledge that is shared and developed through collaborative networks,” highlighting both the outputs of scientific endeavors and the processes involved in their creation~\citep{vicente2018open,leonelli2023philosophy}. Expanding on this perspective, UNESCO defines Open Science as “an inclusive construct that combines various movements and practices aiming to make multilingual scientific knowledge openly available, accessible, and reusable for everyone, to increase scientific collaborations and sharing of information for the benefits of science and society, and to open the processes of scientific knowledge creation, evaluation, and communication to societal actors beyond the traditional scientific community”~\citep{moller2023unesco}.

In practice, Open Science goes beyond providing open access to scientific publications; it also encompasses a wide range of activities, including the sharing of research data, software, and methodologies, all aimed at enhancing transparency and fostering collaboration~\citep{ramachandran_open_2021, mauthner_open_2013}. Among these practices, data sharing has garnered significant attention. Specifically, data sharing refers to the release of data in formats that enable reuse by others~\citep{pasquetto_reuse_2017}. This practice can take many forms, ranging from private exchanges between researchers to more formal mechanisms such as depositing datasets in archives, repositories, domain-specific collections, or library collections. Additionally, researchers may share data by attaching supplemental materials to journal articles or posting datasets on laboratory websites~\citep{wallis_if_2013}.

Data sharing practices offer substantial benefits to both the scientific community and individual researchers. For instance, the free availability of Landsat series data resulted in a twentyfold increase in downloads from the United States Geological Survey between 2009 and 2017, accompanied by a fourfold rise in its use in annual publications~\citep{zhu2019benefits}. This increased accessibility has advanced research applications in land monitoring, enabling studies on surface changes, coastal erosion rates, and glacier fluctuations~\citep{kennedy2014bringing,roy2014landsat,wulder2012opening}.

In addition to broadening research opportunities, data sharing correlates with higher citation rates. For example, \cite{piwowar_data_2013} analyzed 10,555 studies utilizing gene expression microarray data and identified a 9\% citation advantage for papers that shared data. This citation boost varies by discipline. In astronomy, articles linked to open datasets showed a 20\% increase in citation rates~\citep{henneken2011linking}, while paleoclimatology papers with publicly available data experienced a 35\% citation advantage~\citep{sears2011data}. In the social sciences, ~\cite{gleditsch2003posting} examined articles in the Journal of Peace Research and found that those providing data, regardless of format, were cited twice as often as similar articles without accessible data.

Beyond citation impacts, data sharing improves research productivity. A study of over 7,000 NSF and NIH-funded projects found that those with archived data produced a median of 10 publications, compared to only 5 for projects without archived data~\citep{pienta2011enduring}. Additionally, data sharing facilitates peer review and reproducibility, which are essential for verifying research findings and fostering scientific reliability~\citep{peng_reproducible_2011}.

Despite these advantages, there remain open questions on data sharing. One major challenge lies in detecting data-sharing behaviors within publications. Many studies focus on limited datasets or specific disciplines, failing to provide a comprehensive view of data-sharing practices across the scientific community~\citep{zhao_data_2018,koesten_dataset_2020,khan_measuring_2021,stodden_empirical_2018}. For instance, ~\cite{cao_rise_2023} analyzed 1,062,586 arXiv papers in LaTeX format published between 2011 and 2021, but their study focused solely on computer science, physics, and mathematics, leaving other disciplines unexplored. Another limitation arises from the reliance on data availability statements (DAS) as the primary indicator of data sharing~\citep{colavizza_citation_2020,jiao_data_2024,strcic_open_2022}. While useful, DAS are not universally required across fields or journals, creating substantial gaps in understanding the variations in data-sharing practices. Furthermore, as dataset reference standards evolve, mentions of datasets are no longer confined to DAS alone but may appear in other sections of publications~\citep{cao_rise_2023}. 

To address these gaps, our study employs large-scale full-text analysis to investigate data sharing and reuse patterns comprehensively. We aim to answer the following questions:
\begin{itemize}
    \item To what extent is research data released, reused, and referenced across scientific disciplines?
    \item How do releases, reuses, and references change across fields and over time?
\end{itemize}

Our analysis uses the PubMed Open Access (OA) collection, consisting of over 5.7 million full-text articles. To identify the datasets referenced in the publications, we relied on the repository list provided by the European Research Council (ERC) \url{https://zenodo.org/records/7728016}~\citep{jahn_2023_7728016}. This repository encompasses all research funded by the ERC, offering valuable insights into the availability and characteristics of data repositories across diverse research disciplines. 
For their work, the authors considered 220 repositories, identifying 137 trusted data repositories and 74 trusted literature repositories. For our investigation, we rely on the ERC data repositories list to detect and extract mentions to datasets in the full text of papers, as our primary emphasis lies in understanding the availability and nature of repositories across various research fields. Through natural language processing (NLP), this study categorizes data citation intent, such as release, reuse, and reference. This approach provides a nuanced understanding of data citation practices and offers an innovative methodology for analyzing data reuse intentions within scientific literature.

Our findings are expected to shed light on how research data repositories are utilized across diverse scientific fields. By providing insights into data citation patterns, this research aims to guide repository development strategies and contribute to the advancement of open science.

\section{Previous Work}
\label{sec:headings}

\subsection{Open science and research data}
Interest in open science has been growing steadily, with a noticeable increase in the adoption and enforcement of open science practices across disciplines. For instance, funding organizations such as the European Commission require grant recipients to comply with open-access publishing policies under frameworks like Horizon Europe, aiming to enhance the accessibility and dissemination of research outputs to broader audiences~\citep{eu_openscience}. Similarly, numerous academic journals and institutions now mandate practices such as data sharing and methodological transparency as part of their publication and evaluation processes~\citep{robson2021promoting,gorgolewski2016practical}. Moreover, open science communities play a pivotal role in facilitating the large-scale transition of researchers toward open science practices~\citep{armeni2021towards}.

Open science practices extend beyond open-access publishing and include the early sharing of research outputs. For example, platforms like arXiv and bioRxiv enable the dissemination of preprints, fostering early access to findings. Furthermore, open science encourages the public sharing of data and code, often hosted on online repositories such as Zenodo and GitHub, thereby improving research reproducibility and scalability. Open science also promotes rigorous and transparent research design, exemplified by practices like study preregistration~\citep{gopal2018adherence}. 

Substantial evidence indicates that open science practices offer significant advantages over traditional closed practices~\citep{mckiernan_how_2016}. Open-access articles, for example, not only garner broader academic attention and higher citation rates~\citep{huang2024open} but also attract greater engagement from the general public and news media compared to paywalled articles~\citep{schultz2021all,yang2024open}. Furthermore, open science has been shown to accelerate scientific discovery in specific fields~\citep{woelfle2011open}, enhance research transparency, and improve reproducibility~\citep{besanccon2021open}. These benefits play a critical role in addressing challenges associated with the reproducibility crisis~\citep{open2015estimating}.

In today’s data-driven research landscape, the collection, analysis, and interpretation of large datasets are critical to scientific discovery. Among the pillars of open science, research data is particularly vital for promoting transparency and reproducibility. Access to well-documented research data facilitates independent verification of results, supports secondary analyses, and fosters interdisciplinary collaboration, thereby amplifying the impact of scientific inquiry~\citep{hossain2016state,milham_assessment_2018}. There is evidence that integrated data sets have been instrumental in driving biomedical discoveries and drug development~\citep{shahin2020open}.

The advantages of sharing research data are far-reaching, enhancing both the visibility and reuse of research outputs while maximizing the impact of funding agencies' investments~\citep{los_riding_2010}. Recognizing these benefits, governments and funding bodies worldwide have implemented policies to incentivize open data practices. The United States pioneered such efforts as early as 1991~\citep{bromley_policy_1991}, with countries like China, the United Kingdom, and Australia subsequently strengthening their data management frameworks~\citep{china____policy,uk_policy,australia_policy}. In Europe, the Horizon 2020 initiative introduced the Open Research Data Pilot (ODP) to improve data accessibility and establish credibility in data-sharing practices. Leading funding agencies, including the NSF, NIH, and the UK’s Economic and Social Research Council, now require grant applicants to submit data management plans as part of their application process~\citep{smith2012institutional,spengler2012data}. Publishers such as Elsevier, PLOS, Springer, and Nature have also adopted policies that encourage or mandate data citation within reference lists, promoting transparency and accountability in scientific research~\citep{cousijn_data_2018,walton_data_2010,plos_policy,Springer_policy}.

For researchers, open data practices offer additional benefits: they facilitate the development of scientific software~\citep{niemeyer_challenge_2016}, increase research productivity~\citep{mcnaught_changing_2015}, and promote a collaborative data-sharing culture within the scientific community~\citep{belter_measuring_2014}. By aligning incentives for researchers, funders, and publishers, these policies collectively strengthen the foundation for transparent, reproducible, and impactful research.

However, significant barriers continue to hinder the widespread adoption of open data. These include limited incentives, inconsistent citation practices, concerns about data quality, and researchers' reluctance to relinquish control over their data. Additionally, a lack of awareness and insufficient support mechanisms exacerbate these challenges~\citep{chawinga_global_2019,gajbe_evaluation_2021}. Practical issues such as time constraints, inadequate funding, and insufficient institutional support further impede progress~\citep{tenopir_data_2011,tenopir_data_2020}. Deficiencies in archival standards and infrastructure also contribute to low rates of data sharing~\citep{markiewicz_openneuro_2021}. For example, studies that sought to obtain data directly from authors reported low success rates—ranging from 27\% to 59\%, depending on the discipline and geographical context~\citep{tedersoo_data_2021}. Even among papers with data availability statements claiming “data available upon request,” compliance remains low. A 2018 study revealed that only 44\% of authors shared their data when requested~\citep{stodden_empirical_2018}, a finding corroborated by subsequent research~\citep{strcic_open_2022,danchev_evaluation_2021}. 

\subsection{Sharing and reuse of research data}
Research on data sharing and reuse remains in an exploratory stage, with scholars using various data sources and quantitative methods to analyze and discuss data reuse and sharing behaviors in publications.

Disciplinary differences in data citation practices have been a focal point. For instance, ~\cite{park_informal_2018} examined samples from biological and biomedical sciences in the Data Citation Index (DCI), revealing that informal citations within article text are more prevalent than formal citations in reference sections. Similarly, ~\cite{robinson-garcia_analyzing_2016} also utilized DCI data to examine the varying uses of datasets and data studies across disciplines. Their analysis found that datasets were most frequently cited in the fields of science and engineering \& technology, whereas data studies played a more prominent role in the social sciences and arts \& humanities. ~\cite{park_examination_2017} analyzed 148 articles from the Web of Science Data Citation Index to identify factors influencing data sharing and reuse. They found that formal data citation remains relatively uncommon, while informal references in the main text are more typical.

Certain factors have been found to influence researchers' willingness and ability to share and reuse datasets. Studies suggest a correlation between dataset sharing and higher citation rates~\citep{piwowar_data_2013,piwowar_sharing_2007}. Authors also tend to reuse their own shared data, resulting in higher self-citation rates~\citep{robinson-garcia_analyzing_2016}. Data-sharing practices vary notably by discipline, suggesting a need for tailored approaches for each field~\citep{helbig_supporting_2015,torres-salinas_how_2014}. Furthermore, data-sharing rates vary by scientific field~\citep{tenopir_data_2011}, and researchers’ data-sharing behaviors and perceptions differ across age groups and geographical locations~\citep{tenopir_changes_2015}. Certain data types, such as survey, aggregated, and sequence data, receive more frequent citations and higher altmetric scores~\citep{peters_research_2015}.

Studies of data-sharing behavior highlight the impact of shared data on research practices. For instance, an analysis of 600 articles across PLOS journals showed that 74\% of studies rely on datasets created by authors, with fewer reusing prior datasets~\citep{zhao_data_2018}. In biodiversity research, studies using Global Biodiversity Information Facility (GBIF) data demonstrate a rise in open data use, though best practices for data citation remain underutilized~\citep{khan_measuring_2021}.

In addition, journal compliance policies for data sharing have improved, with an increase in the use of repositories instead of supplementary materials for data storage~\citep{jiao_data_2024}. However, data availability statements (DAS) remain inconsistent, especially in COVID-19 research, where only a quarter of preprints provide explicit data-sharing statements~\citep{strcic_open_2022}.

Despite these findings, certain gaps in data-sharing and reuse research remain, particularly in the context of cross-disciplinary data-sharing practices. Most studies are based on samples or case studies from specific fields or repositories, lacking comprehensive cross-disciplinary insights~\citep{kafkas_database_2015, piwowar_beginning_2011, zhao_data_2018,khan_measuring_2021,cao_rise_2023}. Furthermore, the use of data availability statements to accurately identify datasets in academic publications remains limited~\citep{jiao_data_2024,strcic_open_2022}. Some studies suggest that datasets are more commonly cited informally within the text, as opposed to formal citations in references~\citep{belter_measuring_2014,kafkas_database_2015}. While some researchers use the Data Citation Index (DCI) to examine dataset usage, the DCI's focus on natural sciences results in limited coverage across disciplines~\citep{silvello_theory_2017,park_informal_2018,park_examination_2017}, with citation patterns that remain incomplete~\citep{robinson-garcia_analyzing_2016}.

One related study, \cite{cao_rise_2023} investigated the adoption of data and method-sharing practices by analyzing a dataset of 1.1 million arXiv papers, concentrating on physics, mathematics, and computer science. They utilized regular expression matching to extract URLs from the LaTeX-formatted full text of these papers, classifying the URLs as ``data URLs'' or ``method URLs'' using manual annotation and a fine-tuned SciBERT model. Their findings highlighted a growing trend in link-sharing for methods and data, with an increasing number of papers incorporating such URLs over time. They also noted a rise in the reuse of the same links across papers, particularly in computer science, indicating a possible expansion of reproducibility efforts. Furthermore, the analysis revealed a consolidation of links within fewer web domains, such as GitHub, over time. Importantly, papers featuring shared links tended to have a higher citation impact, especially when the links remained active, underscoring the practical benefits of data-sharing practices.

While this study represents a valuable contribution by leveraging full-text analysis on a large dataset, it has notable limitations. Its focus on preprint articles and specific disciplines (physics, mathematics, and computer science) may restrict the generalizability of its findings. Preprints are not universally utilized across all academic disciplines, meaning the dataset may not adequately capture fields where preprint culture is less established. Moreover, the exclusion of formally published articles leaves unanswered questions about potential differences in data-sharing practices between preprints and peer-reviewed publications. Considering the diverse adoption rates of data-sharing practices across scientific disciplines, expanding this research to include formally published articles and additional fields would offer a more comprehensive understanding of how data-sharing practices vary and evolve.

To deepen our understanding of data-sharing and reuse practices, further work across disciplines is essential. Our study seeks to provide a more comprehensive perspective on data use across scientific fields, filling gaps left by previous research that focused on specific disciplines or datasets. By broadening the scope of analysis, the study aims to offer practical insights into the factors influencing data-sharing practices and the variability observed across disciplines. These insights can help foster the adoption of standardized practices and promote a more widespread culture of data-sharing within the research community, ultimately enhancing collaboration, reproducibility, and the overall impact of scientific research.

\section{Methodology}
\label{sec:Methodology}
The data utilized in this study was obtained from the PubMed Open Access collection\footnote{\url{https://www.ncbi.nlm.nih.gov/pmc/tools/openftlist/}} as of March 2024. The total number of publications considered in this dataset is N = 5,704,648 (3,772,464 of oa\_comm,1,502,488 of oa\_noncomm, 429,696 of oa\_other)\footnote{Commercial Use Allowed (oa\_comm): CC0, CC BY, CC BY-SA, and CC BY-ND licenses;
Non-Commercial Use Only (oa\_noncomm): CC BY-NC, CC BY-NC-SA, CC BY-NC-ND;
Other (oa\_other): no machine-readable license, no license, or a custom license. \url{https://pmc.ncbi.nlm.nih.gov/tools/ftp/}}. To enhance the dataset with additional information such as publication dates, citation counts, and disciplines, we queried the Dimensions API (March 2024).

To extract the relevant data repositories mentioned in each paper, we implemented a series of processing steps.

Firstly, we employed regular expression (regex) matching to identify repositories from the full text based on their URLs. Our approach involved applying a unified rule across all URLs to strip away the protocol (http, https) and subdomain (www). For example, from the URL `https://meertens.knaw.nl/en/collections/', we retained `meertens.knaw.nl/en/collections'. Preliminary evidence suggests that this approach enhances resource availability compared to relying solely on data availability statements~\citep{federer_long-term_2022,cao_rise_2023}. The comprehensive list of repository links is available in our repository\footnote{\url{https://github.com/alsowbdxa/Research_Data_in_Scientific_Publications/blob/main/Codes/dataset_urls.xlsx}}. Secondly, to ensure consistency in the repository names or URLs, we converted the entire text of the paper and our domain list to lowercase during the matching process.

\begin{figure}[H]
    \centering
    \includegraphics[width=0.8\linewidth]{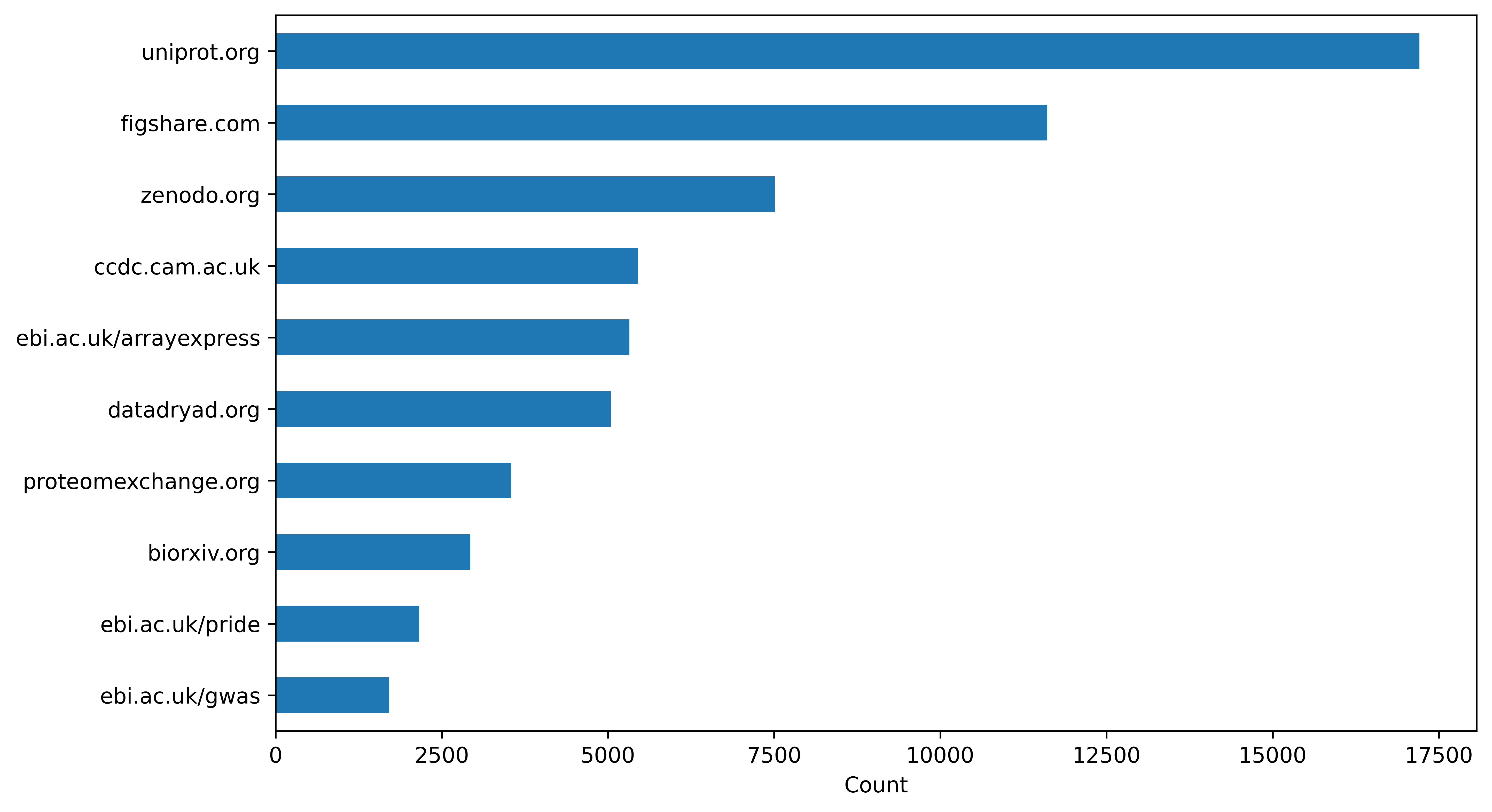}
    \caption{Top 10 repositories by frequency.}
    \label{fig:top10_repo}
\end{figure}

After matching, we successfully extracted 69,090 articles (1.2\%) from the PubMed Open Access collection dataset that included at least one repository link. Figure \ref{fig:top10_repo} illustrates the top 10 repositories appearing in the dataset, ranked by frequency.

Subsequently, these 69,090 papers served as the foundation for constructing our annotation dataset used in model training. Based on our observations and experience, we found that the intention of usage of the repository is typically conveyed within the core sentence (the sentence containing the repository link) and one or two sentences around it. To maximize the information captured while maintaining simplicity, we adopted the same method as a previous study, which incorporates the sentences immediately preceding and following the core sentence to provide additional context ~\citep{zhang2022towards}. Specifically, we retained the two sentences preceding and the two sentences following the core sentence as the context. This context, often regarded as crucial, provides additional information to better identify and classify the intended use of the mentioned datasets~\citep{koesten_dataset_2020}.  Using this method, we processed all the papers and randomly selected 1,000 articles (1.4\% of the 69,090 papers) for annotation.

The manual annotation of these contexts was carried out using Label Studio. We classified the intention of contexts into four categories: 
\begin{itemize}
    \item \textbf{Release}
    The context of the repository mentioned indicates that the paper releases a new dataset on the repository, or generates a dataset by integrating diverse published datasets and releasing it on the repository. A repository ID is typically provided alongside the mention or in the paper.
    
        \begin{quote}
        Example 1: \textit{The datasets are currently for private access during this review period, which can be accessed through: \url{https://datadryad.org/stash/share/yRDf1Kmj9_hR_IIGg_vukBVNUmmB9tm_j8v1BZ721A}.}
        \end{quote}
        
        \begin{quote}
        Example 2: \textit{Raw microarray data have been deposited in compliance to MIAME guidelines at ArrayExpress database (\url{http://www.ebi.ac.uk/arrayexpress}), with accession number E-TABM-1215 release date June 11, 2012. Gene subsets corresponding to each combination of responses analyzed by microarrays were defined from Venn diagrams indicating the number of the included genes.}
        \end{quote}
        
    \item \textbf{Reuse}
    The context surrounding the repository mentioned reveals that the paper directly employed a published dataset hosted on the repository. A repository ID is typically provided alongside the mention or in the paper.
        \begin{quote}
        Example 1: \textit{The European Commission do not accept any responsibility for use that may be made of the information it contains. All data used in the current study is publicly available. Summary statistics for IBS can be download from European Bioinformatics Institute GWAS Catalog (\url{https://www.ebi.ac.uk/gwas/)}. Summary statistics for neuroticism can be downloaded from \url{https://ctg.cncr.nl/software/summary\_statistics} and \url{http://www.ccace.ed.ac.uk}. Summary statistics for depression can be downloaded from \url{https://datashare.ed.ac.uk/handle/10283/3203}.}
        \end{quote}

        \begin{quote}
        Example 2: \textit{The land use data were obtained from the 30-m annual land cover datasets and its dynamics in China from 1990 to 2020 (\url{https://zenodo.org/record/5210928\#.Y9TDU3ZBxD}).}
        \end{quote}

    \item \textbf{Reference}
    The context surrounding the repository mentioned indicates that the paper references the repository, possibly to compare different datasets or for context. Importantly, the authors' work is not reliant on this dataset, nor have they produced a new dataset based on it.
        \begin{quote}    
        Example 1: \textit{Furthermore, in Table S3, we also list the top 20 ranked potential phosphorylation sites for MAPK1, in which Tyr325 and Tyr331 of FOS (P01100) has been confirmed to be modified by this kinase (\url{http://www.uniprot.org/uniprot/P01100\#ptm\_processing}).}
        \end{quote}

        \begin{quote}
        Example 2: \textit{Some of the resources used an ontology, e.g., Disease Ontology, a taxonomy such as MeSH [24], or cross-referenced another resource such as OMIM. Diseases and phenotypes are often mixed in the same resource and sometimes in the same category annotation. For example, the European Variation Archive (\url{EVA – http://www.ebi.ac.uk/eva/}) [25] trait names’ labeling uses a mixed set of vocabularies from HP, SNOMED-CT, OMIM, and non-standardized local identifiers used internally at source from the ClinVar records. The identifiers of the record’s cross-references for each trait name are not equivalently represented - e.g., trait name ‘congenital adrenal hyperplasia’ in EVA contains identifiers for SNOMED-CT, HP, but not for OMIM. This trait name also links to a non-standardized internal identifier used at the Office of Rare Disease.}
        \end{quote}

    \item \textbf{Nothing}
    Occasionally, we encountered erroneous or non-related hits. While using repository links to identify repositories within the full text, we found that sometimes these links did not solely indicate repositories but could also convey other meanings. In such cases, we labelled it as `Nothing.'

        \begin{quote}
        Example 1: \textit{The following link will take you to the Dryad record for your article, so you won't have to re-enter its bibliographic information, and can upload your files directly: \url{http://datadryad.org/submit?journalID=pgenetics\&manu=PGENETICS-D-19-01831R2} More information about depositing data in Dryad is available at \url{http://www.datadryad.org/depositing}.}
        \end{quote}

        \begin{quote}
        Example 2: \textit{Assessing the impact of autologous neutralizing antibodies on viral rebound in postnatally SHIV-infected ART-treated infant rhesus macaques  14 9 2023 2023.07.22.550159 \url{http://biorxiv.org/lookup/doi/10.1101/2023.07.22.550159} Abstract  While the benefits of early antiretroviral therapy (ART) initiation in perinatally infected infants are well documented, early ART initiation is not always possible in postnatal pediatric HIV infections, which account for the majority of pediatric HIV cases worldwide. The timing of onset of ART initiation is likely to affect the size of the latent viral reservoir established, as well as the development of adaptive immune responses, such as the generation of neutralizing antibody responses against the virus.}
        \end{quote}
    
\end{itemize}

Following the definition of these four intentions, we manually annotate the annotation subset (1,000 articles with 1328 contexts), and we get 670 contexts with the label `Release,' 119 contexts with `Reference,' 453 contexts with `Reuse' and 86 contexts with `Nothing.' Then we use this subset to train the model.

For model training, we utilized pre-trained models from Hugging Face,  specifically BertForSequenceClassification `bert-base-uncased' for BERT and RobertaForSequenceClassification `roberta-base' for RoBERTa, this model has been validated as delivering optimal performance in most NLP tasks~\citep{devlin2018bert}. Before training, we mapped the original labels to distinct integers, assigning `Release' as label 0, `Reuse' as label 1, `Reference' as label 2, and `Nothing' as label 3. The dataset was then partitioned into training, testing, and validation subsets in an 80-10-10\% split.

To prepare the textual data for modeling, we performed tokenization using BERT and RoBERTa tokenizers respectively, considering a maximum sequence length of 512 tokens, for each sentence. If the total number of tokens is less than 512 (the model's maximum limit), the entire sentence is retained. However, if it exceeds 512 tokens, we employ four different truncation methods:
\begin{itemize}
    \item Method 1
        If it exceeds 512 tokens, we truncate it by retaining the first 512 tokens.
    \item Method 2
        If it exceeds 512 tokens, we truncate it by preserving the last 512 tokens.
    \item Method 3
         If it exceeds 512 tokens, we truncate it by keeping the central 512 tokens. For instance, if it has 1000 tokens, we remove the first 244 and last 244 tokens.
    \item Method 4
         If it exceeds 512 tokens, we truncate it by keeping the first 256 tokens and the last 256 tokens.
\end{itemize}

For each model, we employ each truncation method and evaluate the model based on the F1 score. Additionally, we incorporate the early stopping mechanism in the training process. Specifically, if the F1 score on the validation set shows no improvement over 10 epochs, and the model’s performance starts to degrade, we terminate the training

We present an overview of the performance results achieved by various methods when applied to either the RoBERTa and BERT models\ref{tab:Performance of models by methods}. We see that RoBERTa, specifically when employed with method 2, outperforms the other configurations, boasting a F1 score of 0.902. Building upon this outcome, we further enhance the model's efficacy by merging the training and test subsets. Leveraging the RoBERTa model in conjunction with truncation method 2, we conduct fine-tuning to optimize its performance. The resultant refined model is subsequently subjected to testing on the validation dataset.

\begin{table}[H]
\centering
\resizebox{0.7\textwidth}{!}{%
\begin{tabular}{|c|c|c|c|c|c|}
\hline
          &          & Accuracy & Precision & Recall  & F1      \\ \hline
\multirowcell{4}{RoBERTa}   & Method 1 & 0.896             & 0.897              & 0.896           & 0.894 \\
\cline{2-6}
          & Method 2 & \textbf{0.902}             & \textbf{0.902}              & \textbf{0.902}          &\textbf{0.902} \\
\cline{2-6}          & Method 3 & 0.886             & 0.885              & 0.886           & 0.885 \\
\cline{2-6}          & Method 4  & 0.885             & 0.890              & 0.890           & 0.886 \\ 
\hline     \cline{2-6}      
\multirowcell{4}{BERT}      & Method 1 & 0.876             & 0.882              & 0.876           & 0.876 \\
\cline{2-6} 
          & Method 2 & 0.876             & 0.874              & 0.876           & 0.874 \\
\cline{2-6}          & Method 3 & 0.870             & 0.873              & 0.870           & 0.871 \\
\cline{2-6}          & Method 4 & 0.855             & 0.864              & 0.855           & 0.857  \\ \hline
\end{tabular}%
}
\caption{Performance of models by methods}
\label{tab:Performance of models by methods}
\end{table}

\begin{figure}[H]
    \centering
    \includegraphics[width=0.8\linewidth]{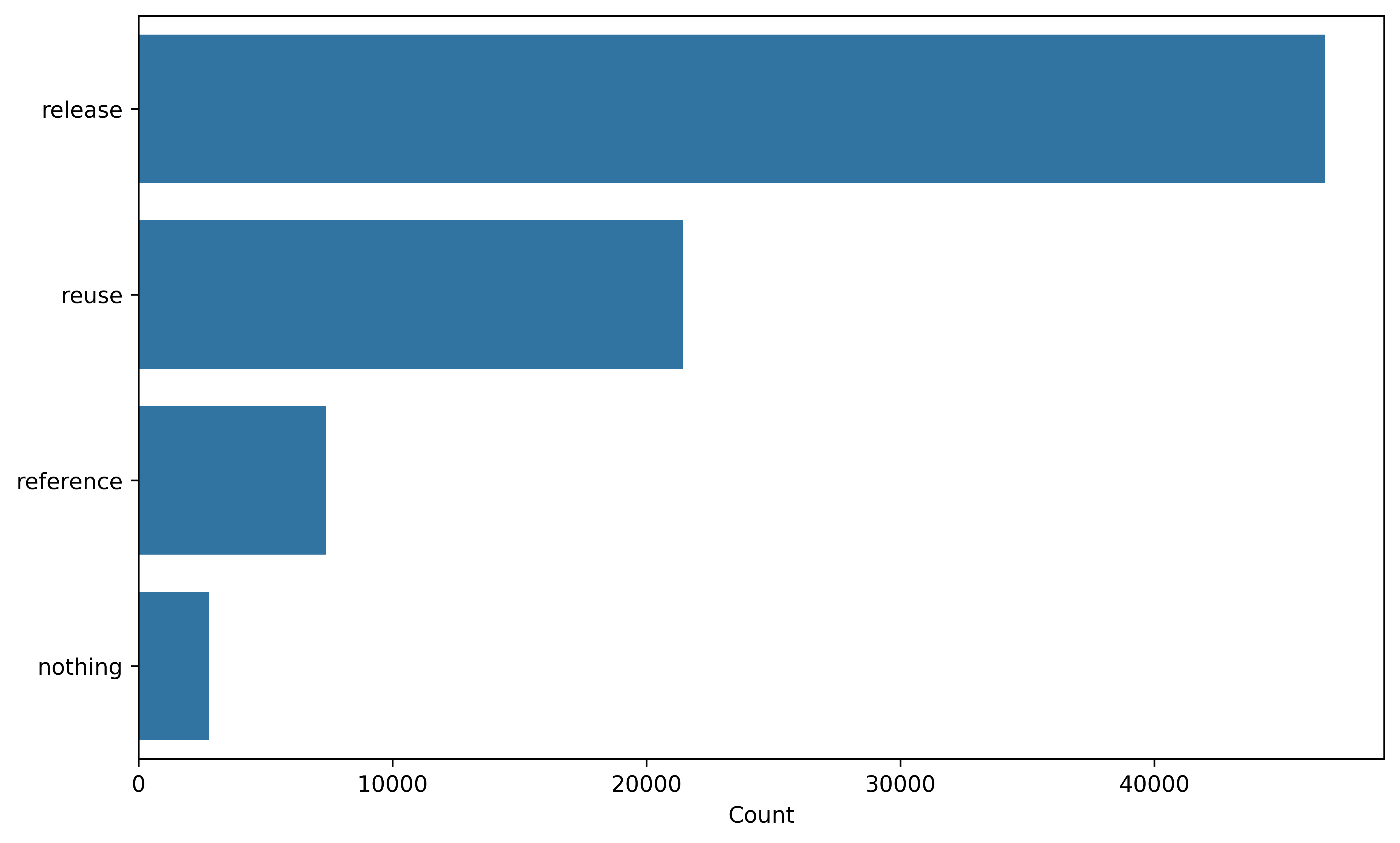}
    \caption{Predicted label distribution}
    \label{fig:predicted_label_distribution}
\end{figure}

Following the training phase, we deploy the trained model to predict the intention for each context within the entire dataset (69,090 articles with 92,267 contexts). The distribution of intentions across the entire dataset is illustrated in Figure \ref{fig:predicted_label_distribution}. 

Specifically, we observe that 55,680 contexts (60.3\%) are labeled as `Release,' 24,809 contexts (26.9\%) as `Reuse,' 8,597 contexts (9.3\%) as `Reference,' and 3,181 contexts (3.5\%) as `Nothing.'
    
We visualize the top 20 repositories along with their respective usage intentions in the figure below:
\begin{figure}[H]
    \centering
    \includegraphics[width=0.8\linewidth]{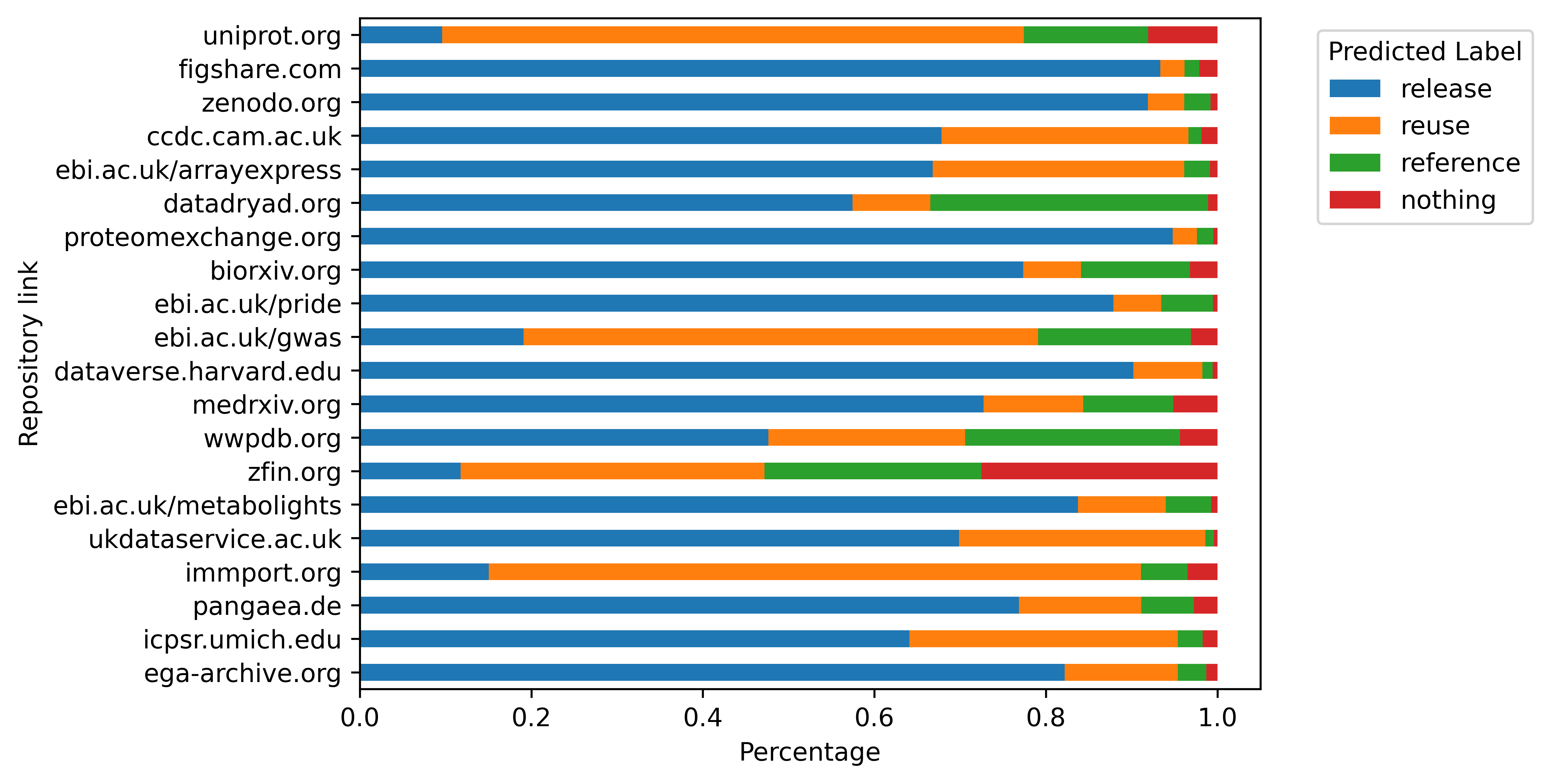}
    \caption{Distribution of intentions by repository.}
    \label{fig: Distribution of using intentions by keyword}
\end{figure}

The patterns of using intentions across repositories are highlighted in Figure \ref{fig: Distribution of using intentions by keyword}. The figure further validates the accuracy of our classification to some extent. Repositories such as Figshare and Zenodo, often utilized for publishing datasets, demonstrate a higher frequency of the `Release' type. Conversely, repositories like Uniport\footnote{A prominent free-access collection of protein sequences and their annotations, supporting fields like biology, medicine, and biotechnology. \url{https://pmc.ncbi.nlm.nih.gov/articles/PMC4384041/}} and ebi.ac.uk\footnote{The world’s most comprehensive suite of freely available data resources and tools for life science research. \url{https://www.ebi.ac.uk/about}}, dedicated to supplying datasets for research analysis, display a predilection for the `Reuse' type.

\section{Results}
We start by providing a description of our findings, followed by an in-depth discussion in the subsequent section.

\begin{figure}[htpb]
    \centering
    \includegraphics[width=0.8\linewidth]{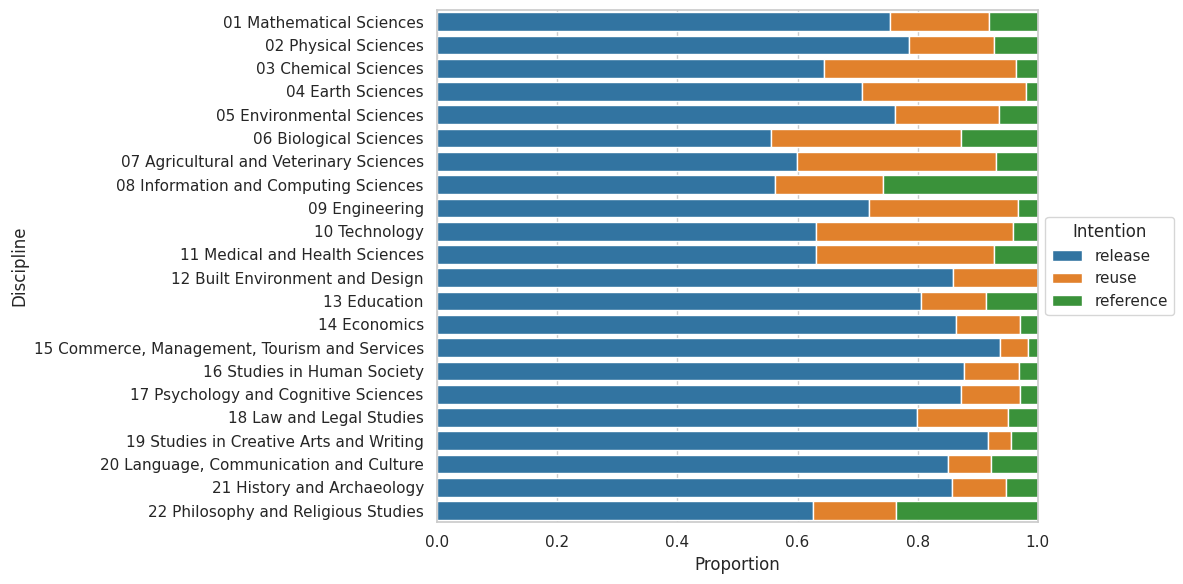}
    \caption{Distribution of intentions across disciplines.}
    \label{fig:discipline2intention}
\end{figure}

Figure \ref{fig:discipline2intention} presents the proportional distribution of three intentions across various academic disciplines. The horizontal bar plot includes 22 disciplines, each represented on the y-axis, while the x-axis shows the proportion of each intention relative to the total papers for that field, which is calculated based on fractional counting. 

In most disciplines, the `release' intention dominates, indicating a strong preference for openly sharing data. Specifically, the top five disciplines for releasing datasets are `Commerce, Management, Tourism and Services', `Studies in Creative Arts and Writing', `Studies in Human Society', `Psychology and Cognitive Sciences' and `Economics'. Conversely, the proportions of released datasets are lowest in `Biological Sciences', `Information and Computing Sciences' and `Agricultural and Veterinary Sciences'.

Regarding the intention of reuse, STEM-related fields generally exhibit a higher proportion of reuse. Notably, `Agricultural and Veterinary Sciences', `Technology', `Chemical Sciences', `Biological Sciences', `Medical and Health Sciences' have over 30\% of mentions indicating dataset reuse.

For reference intention, datasets are referenced less frequently across all disciplines, with two exceptions: `Information and Computing Sciences' and `Philosophy and Religious Studies'.

\begin{figure}[H]
    \centering
    \includegraphics[width=0.8\linewidth]{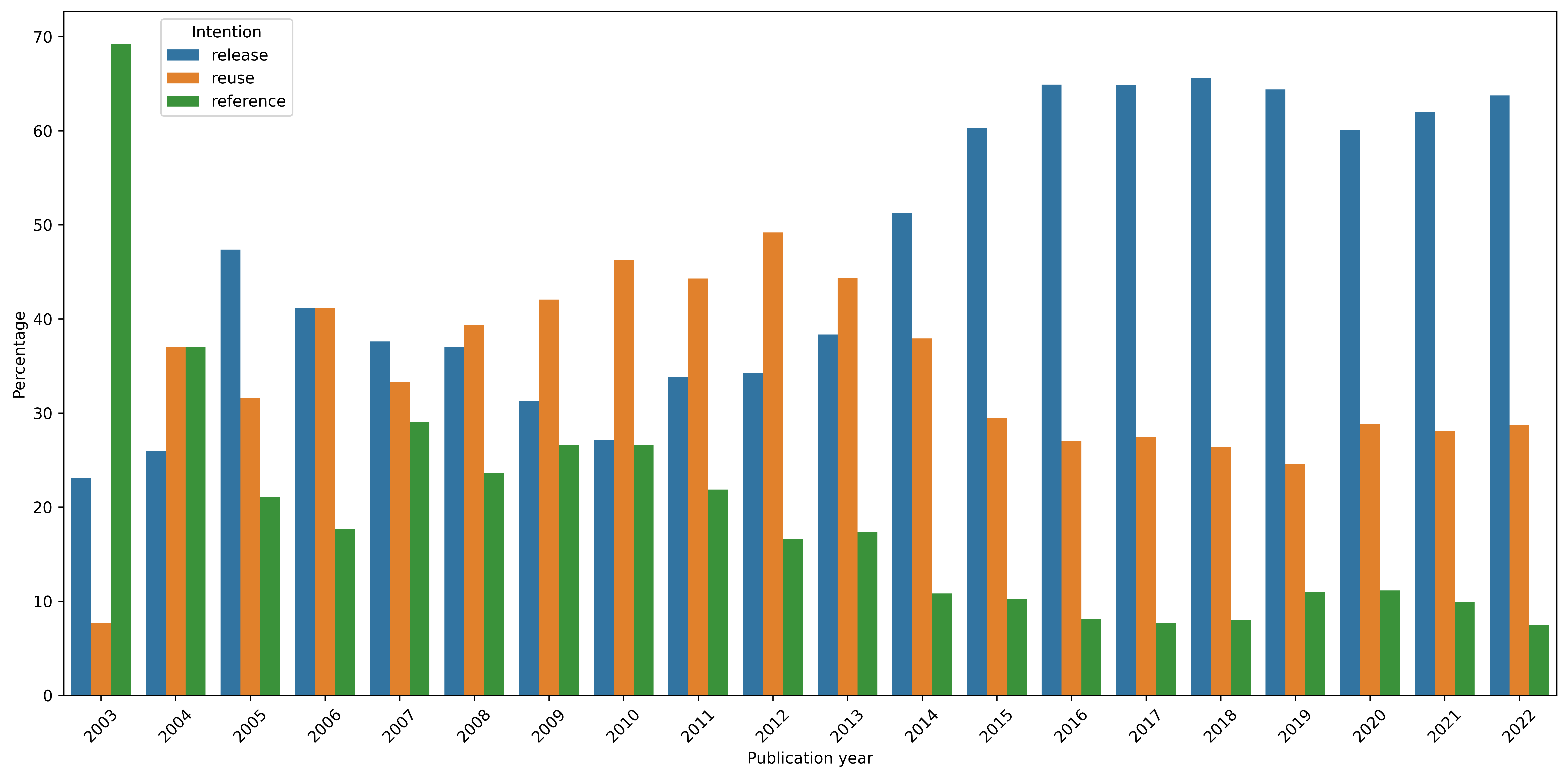}
    \caption{Distribution of intention over time.}
    \label{fig:Distribution of Intention Over Time}
\end{figure}

Furthermore, figure \ref{fig:Distribution of Intention Over Time} shows the distribution of different mention intentions in the dataset (`reuse,' `release,' and `reference') over time, with the x-axis representing the years and the y-axis showing the percentage of each intention.

The figure reveals trends in dataset usage across publications. From 2007 to 2012, the intention to reuse datasets increased, while the intention to release datasets remained relatively low and even declined slightly. However, starting in 2012, the trends shifted. The intention to release datasets sharply increased and consistently remained high (around 60\%), while the intention to reuse datasets decreased significantly, dropping from 50\% to approximately 30\%.

To explore the disciplinary interconnections in the contexts of release, reuse, and reference intentions, we constructed co-occurrence networks for each intention via VOSviewer. Specifically, a publication may encompass multiple disciplines, the co-occurrence of macro disciplines associated with that publication is treated as a co-occurrence instance. The co-occurrence distribution for each intention network is subsequently calculated using fractional counting. In these networks, nodes represent disciplines, edges represent the co-occurrence of two disciplines within the same article, and the color means the cluster identified based the VOSviewer\citep{van2010software}. The strength of connections is determined by the frequency of such co-occurrences.

\begin{figure}[H]
    \centering
    \includegraphics[width=0.6\linewidth]{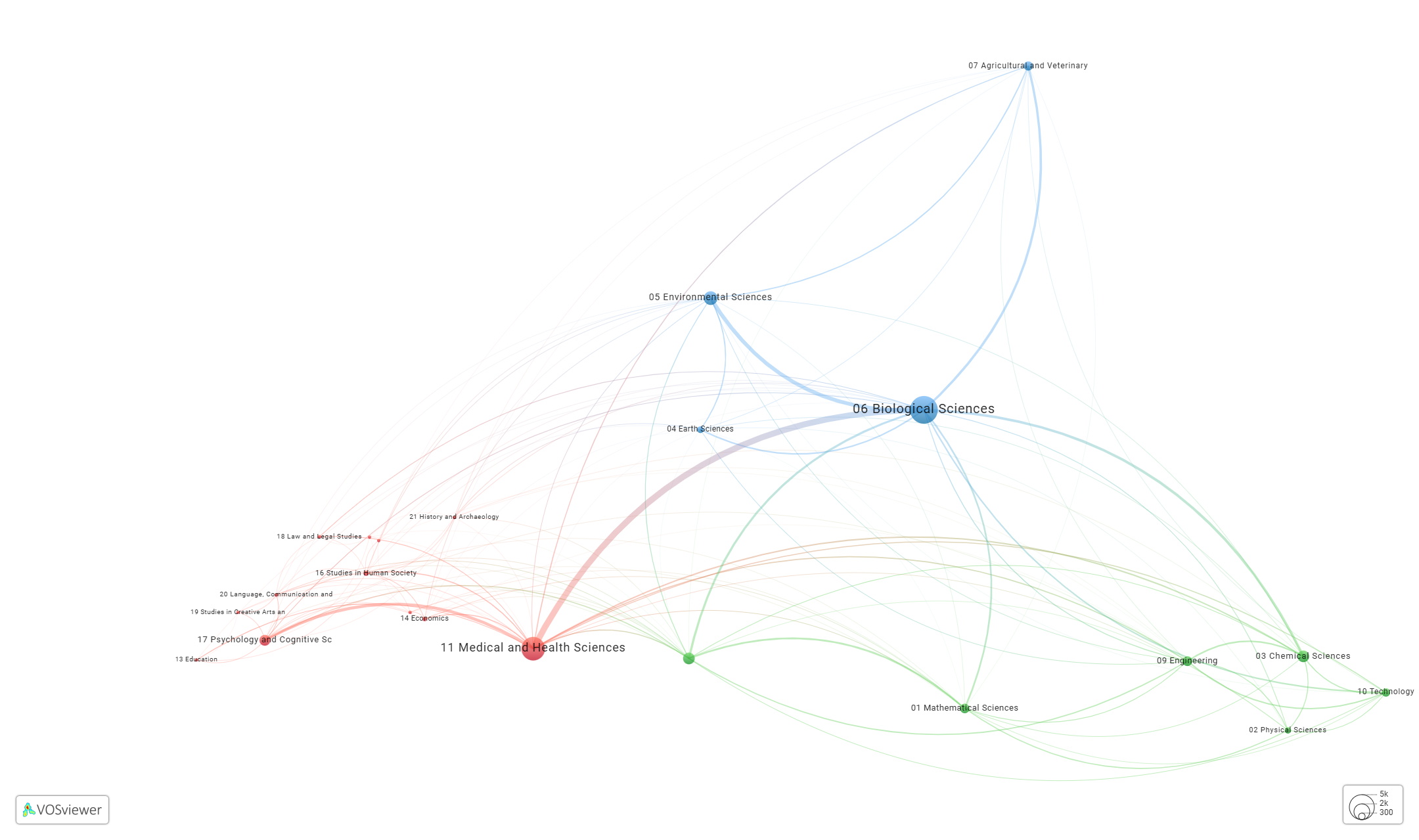}
    \caption{Co-occurrence network of release}
    \label{fig:co-occuarance_network_release}
\end{figure}

The co-occurrence network for the release intention (Figure \ref{fig:co-occuarance_network_release}) underscores the pivotal role of Biological Sciences and Medical and Health Sciences, which form strong connections with related disciplines such as Agricultural and Veterinary Sciences, Environmental Sciences, and Chemical Sciences. 

\begin{figure}[H]
    \centering
    \includegraphics[width=0.6\linewidth]{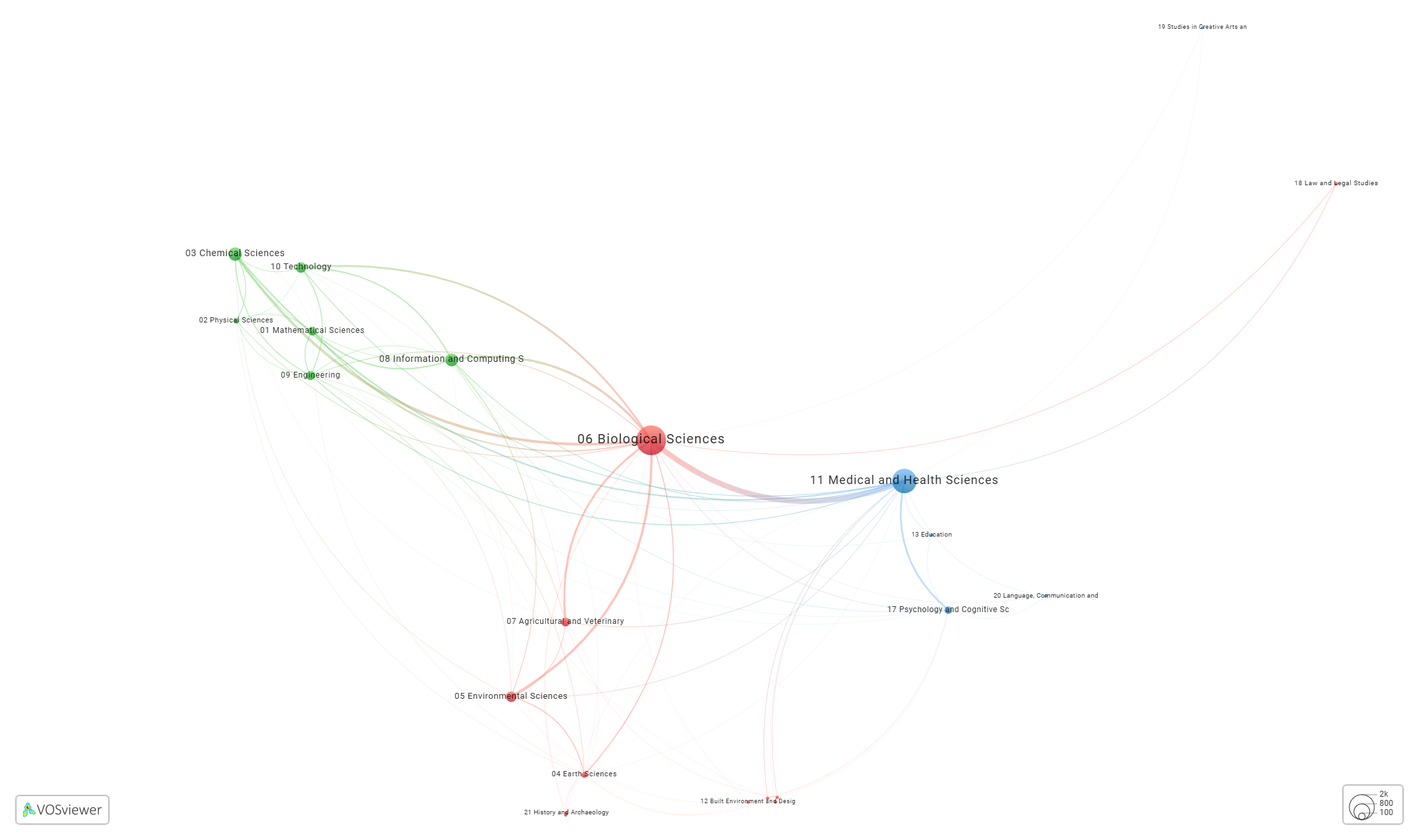}
    \caption{Co-occurrence network of reuse}
    \label{fig:co-occuarance_network_reuse}
\end{figure}

For the reuse intention (Figure \ref{fig:co-occuarance_network_reuse}), the network displays a more dispersed pattern of connections. While Biological Sciences and Medical and Health Sciences remain central, the network highlights strong links with Information and Computing Sciences and Mathematical Sciences, illustrating the growing importance of computational and data-driven approaches in reused research. Additionally, significant connections to social science disciplines, such as Studies in Human Society and Education, suggest that data reuse is becoming increasingly relevant across diverse academic contexts.

\begin{figure}[H]
    \centering
    \includegraphics[width=0.6\linewidth]{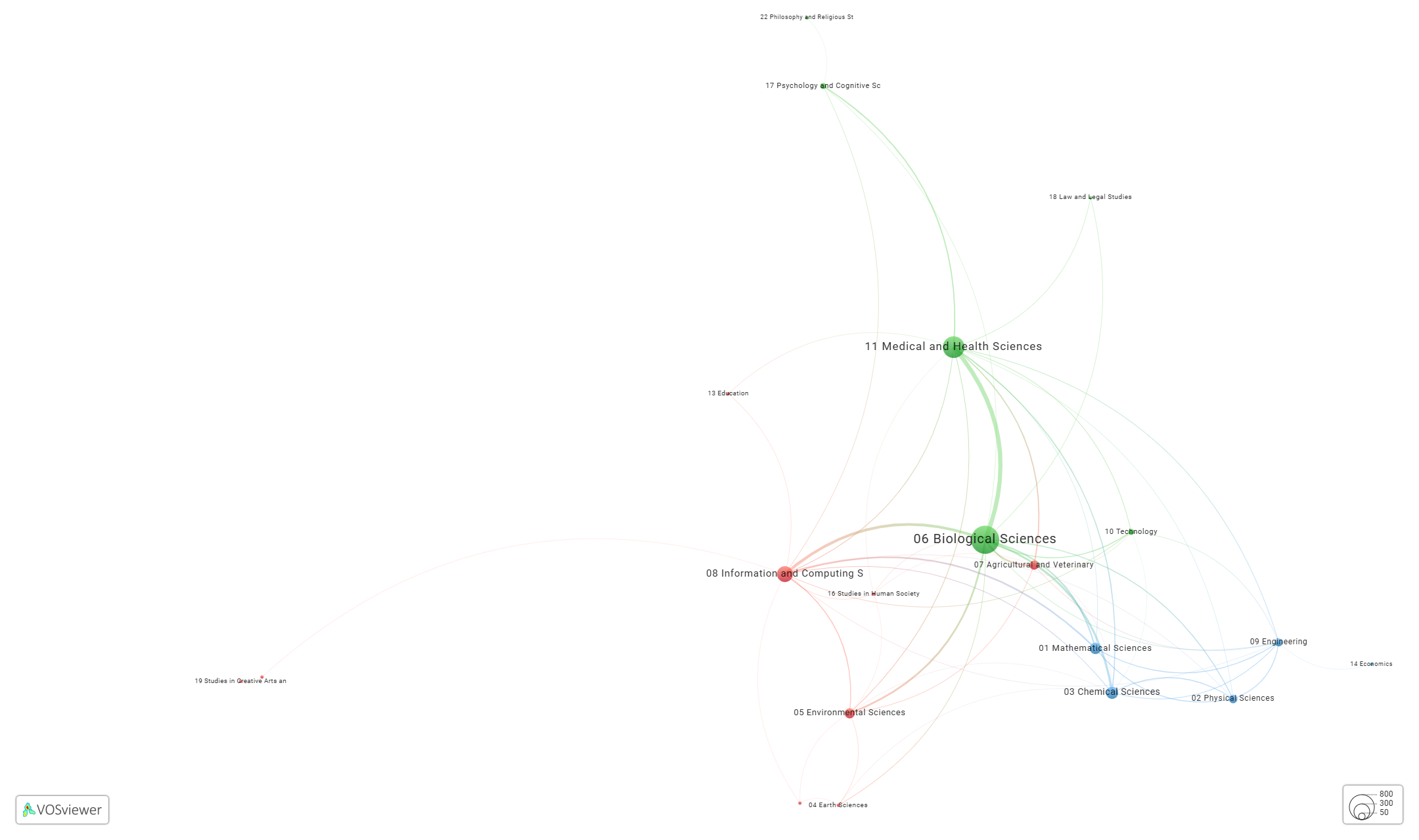}
    \caption{Co-occurrence network of reference}
    \label{fig:co-occuarance_network_reference}
\end{figure}

The reference intention network (Figure \ref{fig:co-occuarance_network_reference}) exhibits a distinct structure, with Biological Sciences maintaining a central role but with more dispersed connections compared to the other two intentions.

\section{Discussion}
\subsection{Extent of data release, reuse, and reference across disciplines}

Our analysis reveals that the release intention is the dominant mode of data sharing across most disciplines. This prevalence is likely driven by the widespread influence of open science policies, the growing emphasis on transparency and reproducibility, and the expanding availability of data repositories. Public repositories such as Zenodo and Figshare, as well as discipline-specific platforms like UniProt and CCDC, have made data sharing more accessible. Moreover, the relatively lower technical and cognitive barriers to releasing data, compared to reusing or citing datasets, further contribute to its widespread adoption. This trend is particularly pronounced in disciplines such as `Commerce, Management, Tourism and Services', `Studies in Creative Arts and Writing', `Studies in Human Society', `Psychology and Cognitive Sciences' and `Economics', where data release accounts for over 80\% of all mentions. Such dominance suggests a strong alignment with open science principles in these disciplines. Previous research similarly highlights the prominent role of data studies in the social sciences and arts and humanities ~\citep{jimenez2015analyzing}. These findings are consistent with studies emphasizing the positive impact of open science in fostering transparency, reproducibility, and collaborative research ~\citep{tenopir_changes_2015}.

Conversely, disciplines such as `Biological Sciences', `Information and Computing Sciences' and `Agricultural and Veterinary Sciences' exhibit lower levels of data release, likely due to field-specific challenges, including ethical considerations, sensitivity of data, and proprietary restrictions. Prior studies underscore these barriers, highlighting the complexities associated with consent, privacy, and intellectual property in these fields ~\citep{tenopir_changes_2015, huang2012willing, oushy2015share}. Addressing these challenges requires robust infrastructure and governance mechanisms to ensure compliance. Thus, while the overall trend supports open science initiatives, significant variation persists across disciplines due to these domain-specific barriers and norms.

In contrast, the reuse intention is more prevalent in STEM-related fields such as `Agricultural and Veterinary Sciences', `Technology', `Chemical Sciences', `Biological Sciences', `Medical and Health Sciences', where over 30\% of mentions involve the reuse of existing datasets. This prevalence can be attributed to the availability of shared databases, such as UniProt and CCDC, and the methodological reliance on pre-existing data in these disciplines ~\citep{jimenez2015analyzing}. However, the relatively low reuse proportions in humanities and social sciences suggest that data reuse practices are less institutionalized in these fields. This discrepancy likely arises from variations in data availability, research methodologies, and the perceived value of reusing datasets ~\citep{kim2017scientists}.

The reference intention remains consistently low across most disciplines, with notable exceptions in `Information and Computing Sciences' and `Philosophy and Religious Studies'. This overall low level of dataset referencing highlights a critical issue in academic publishing: datasets are not yet widely recognized as formal research outputs in many fields ~\citep{silvello_theory_2017}. While dataset citation practices are gaining traction in Information and Computing Sciences ~\citep{force2014force}, other disciplines lag behind due to a lack of standardized citation practices and limited awareness of the benefits of dataset citation ~\citep{kratz2014data}. As previous studies suggest, data citation not only provides credit to data creators but also enhances transparency and reproducibility, underscoring its significance in advancing open science ~\citep{altman2015introduction, piwowar_data_2013}.

In addition, Biological Sciences and Medical and Health Sciences play a pivotal role in the connection among disciplines, which consistently serve as hubs across all intentions. These fields highlight the inherently interdisciplinary nature of research, particularly in data release practices, as evidenced by strong connections within the life and health sciences cluster. This observation aligns with previous findings that 86\% of research data published in biological sciences journals are cited by articles from disciplines outside the biological sciences domain in the Web of Science ~\citep{park2022interdisciplinarity}.

\subsection{Temporal trends in releases, reuses, and references}

Temporal trends in data-sharing practices show a clear evolution. From 2007 to 2012, the rise in reuse intentions marked the early stages of data-sharing adoption, driven by large-scale repositories and a growing emphasis on data-driven research ~\citep{tenopir_changes_2015}. The post-2012 surge in data release intentions may align with global open science initiatives, such as the U.S. White House memorandum in 2013, which required federal agencies to increase public access to research results, and the Plan S initiative in 2018, which set standards for immediate open access publication across Europe ~\citep{holdren2013memorandum, plans2018}. These policies have had a substantial societal impact, making millions of academic publications freely accessible to the public and fostering a shift toward collaborative, open science.

Since 2018, all disciplines have seen significant growth in data release activity, with STEM fields showing steady growth in reuse and reference intentions. This signals the increasing normalization of data-driven research practices. However, the delayed adoption of these practices in the humanities suggests ongoing cultural and infrastructural shifts, compounded by challenges such as non-standardized data formats and discipline-specific attitudes toward open science ~\citep{fuhr2021digital}.

Despite the success of open access policies, the decline in reuse intentions highlights ongoing challenges in data discoverability, compatibility, and a lack of incentives for reuse. Issues such as insufficient metadata, unclear licensing terms, and the technical complexity of integrating datasets may continue to hinder effective reuse in new research contexts ~\citep{borgman_conundrum_2012, mayernik2017open}.

We acknowledge certain limitations in our study. First, while we attempted to match dataset mentions in the full text with the repository list provided by the European Research Council, some dataset mentions may have been missed. These could include instances where datasets were not associated with URLs or were not included in the repository list. Additionally, our analysis relies on full-text data, and although we worked with a substantial corpus of millions of publications across disciplines, it should be noted that the dataset mostly represents biomedical and life sciences journal literature. As a result, our findings may not fully capture data-sharing practices in fields where such datasets are less prevalent or not yet widely integrated into the research ecosystem. 

\section{Conclusion}
This study highlights the evolving landscape of data-sharing practices across disciplines, focusing on the intentions of data release, reuse, and reference. The findings indicate that the release intention is the dominant mode of data sharing, with notable variations across disciplines. Fields such as Commerce, Management, and Creative Arts show high levels of data release, reflecting a strong alignment with open science principles, while disciplines like Biological Sciences and Agricultural Sciences face unique challenges related to ethical, legal, and privacy concerns. The reuse intention is particularly prevalent in STEM-related disciplines, emphasizing the growing reliance on shared datasets and computational methodologies in research. However, humanities and social sciences show a delayed adoption of data reuse practices, likely due to factors like limited data availability and infrastructure.

The analysis also reveals a low proportion of dataset referencing across most fields, suggesting that datasets are not yet fully recognized as formal research outputs, despite their increasing role in the research process. Temporal trends indicate that recent open science initiatives have accelerated data release practices, particularly post-2012, yet challenges persist, especially in terms of data discoverability and compatibility for reuse. The findings further highlight the central role of the Biological and Medical Sciences in fostering interdisciplinary data sharing.

This study provides a comprehensive understanding of how data is utilized across different scientific disciplines and offers valuable insights to help institutions and publishers develop better data policies. By identifying trends in data release, reuse, and citation, it can inform strategies to enhance data sharing practices and improve the accessibility and discoverability of datasets. These findings will assist in creating more effective support systems for researchers and encourage broader adoption of open science practices across various fields.








\bibliographystyle{unsrtnat}
\bibliography{references}  

\begin{thebibliography}{92}
\providecommand{\natexlab}[1]{#1}
\providecommand{\url}[1]{\texttt{#1}}
\expandafter\ifx\csname urlstyle\endcsname\relax
  \providecommand{\doi}[1]{doi: #1}\else
  \providecommand{\doi}{doi: \begingroup \urlstyle{rm}\Url}\fi

\bibitem[Willinsky(2005)]{willinsky2005unacknowledged}
John Willinsky.
\newblock The unacknowledged convergence of open source, open access, and open science.
\newblock \emph{First Monday}, 2005.

\bibitem[Moore(2017)]{moore2017genealogy}
Samuel~A Moore.
\newblock A genealogy of open access: negotiations between openness and access to research.
\newblock \emph{Revue fran{\c{c}}aise des sciences de l’information et de la communication}, 11\penalty0 (2), 2017.

\bibitem[Vicente-Saez and Martinez-Fuentes(2018)]{vicente2018open}
Ruben Vicente-Saez and Clara Martinez-Fuentes.
\newblock Open science now: A systematic literature review for an integrated definition.
\newblock \emph{Journal of business research}, 88:\penalty0 428--436, 2018.

\bibitem[Leonelli(2023)]{leonelli2023philosophy}
Sabina Leonelli.
\newblock \emph{Philosophy of open science}.
\newblock Cambridge University Press, 2023.

\bibitem[M{\"o}ller(2023)]{moller2023unesco}
Lutz M{\"o}ller.
\newblock Unesco recommendation on open science.
\newblock In \emph{66. Helmholtz Open Science online seminar}, 2023.

\bibitem[Ramachandran et~al.(2021)Ramachandran, Bugbee, and Murphy]{ramachandran_open_2021}
Rahul Ramachandran, Kaylin Bugbee, and Kevin Murphy.
\newblock From {Open} {Data} to {Open} {Science}.
\newblock \emph{Earth and Space Science}, 8\penalty0 (5):\penalty0 e2020EA001562, May 2021.
\newblock ISSN 2333-5084.
\newblock \doi{10.1029/2020EA001562}.
\newblock URL \url{https://agupubs.onlinelibrary.wiley.com/doi/10.1029/2020EA001562}.

\bibitem[Mauthner and Parry(2013)]{mauthner_open_2013}
Natasha~Susan Mauthner and Odette Parry.
\newblock Open {Access} {Digital} {Data} {Sharing}: {Principles}, {Policies} and {Practices}☆.
\newblock \emph{Social Epistemology}, 27\penalty0 (1):\penalty0 47--67, January 2013.
\newblock ISSN 0269-1728.
\newblock \doi{10.1080/02691728.2012.760663}.
\newblock URL \url{http://www.tandfonline.com/doi/abs/10.1080/02691728.2012.760663}.

\bibitem[Pasquetto et~al.(2017)Pasquetto, Randles, and Borgman]{pasquetto_reuse_2017}
Irene~V. Pasquetto, Bernadette~M. Randles, and Christine~L. Borgman.
\newblock On the {Reuse} of {Scientific} {Data}.
\newblock \emph{Data Science Journal}, 16:\penalty0 8, March 2017.
\newblock ISSN 1683-1470.
\newblock \doi{10.5334/dsj-2017-008}.
\newblock URL \url{https://datascience.codata.org/articles/10.5334/dsj-2017-008}.

\bibitem[Wallis et~al.(2013)Wallis, Rolando, and Borgman]{wallis_if_2013}
Jillian~C. Wallis, Elizabeth Rolando, and Christine~L. Borgman.
\newblock If {We} {Share} {Data}, {Will} {Anyone} {Use} {Them}? {Data} {Sharing} and {Reuse} in the {Long} {Tail} of {Science} and {Technology}.
\newblock \emph{PLOS ONE}, 8\penalty0 (7):\penalty0 e67332, July 2013.
\newblock ISSN 1932-6203.
\newblock \doi{10.1371/journal.pone.0067332}.
\newblock URL \url{https://journals.plos.org/plosone/article?id=10.1371/journal.pone.0067332}.

\bibitem[Zhu et~al.(2019)Zhu, Wulder, Roy, Woodcock, Hansen, Radeloff, Healey, Schaaf, Hostert, Strobl, et~al.]{zhu2019benefits}
Zhe Zhu, Michael~A Wulder, David~P Roy, Curtis~E Woodcock, Matthew~C Hansen, Volker~C Radeloff, Sean~P Healey, Crystal Schaaf, Patrick Hostert, Peter Strobl, et~al.
\newblock Benefits of the free and open landsat data policy.
\newblock \emph{Remote Sensing of Environment}, 224:\penalty0 382--385, 2019.

\bibitem[Kennedy et~al.(2014)Kennedy, Andr{\'e}fou{\"e}t, Cohen, G{\'o}mez, Griffiths, Hais, Healey, Helmer, Hostert, Lyons, et~al.]{kennedy2014bringing}
Robert~E Kennedy, Serge Andr{\'e}fou{\"e}t, Warren~B Cohen, Cristina G{\'o}mez, Patrick Griffiths, Martin Hais, Sean~P Healey, Eileen~H Helmer, Patrick Hostert, Mitchell~B Lyons, et~al.
\newblock Bringing an ecological view of change to landsat-based remote sensing.
\newblock \emph{Frontiers in Ecology and the Environment}, 12\penalty0 (6):\penalty0 339--346, 2014.

\bibitem[Roy et~al.(2014)Roy, Wulder, Loveland, Woodcock, Allen, Anderson, Helder, Irons, Johnson, Kennedy, et~al.]{roy2014landsat}
David~P Roy, Michael~A Wulder, Thomas~R Loveland, Curtis~E Woodcock, Richard~G Allen, Martha~C Anderson, Dennis Helder, James~R Irons, David~M Johnson, Robert Kennedy, et~al.
\newblock Landsat-8: Science and product vision for terrestrial global change research.
\newblock \emph{Remote sensing of Environment}, 145:\penalty0 154--172, 2014.

\bibitem[Wulder et~al.(2012)Wulder, Masek, Cohen, Loveland, and Woodcock]{wulder2012opening}
Michael~A Wulder, Jeffrey~G Masek, Warren~B Cohen, Thomas~R Loveland, and Curtis~E Woodcock.
\newblock Opening the archive: How free data has enabled the science and monitoring promise of landsat.
\newblock \emph{Remote Sensing of Environment}, 122:\penalty0 2--10, 2012.

\bibitem[Piwowar and Vision(2013)]{piwowar_data_2013}
Heather~A. Piwowar and Todd~J. Vision.
\newblock Data reuse and the open data citation advantage.
\newblock \emph{PeerJ}, 1:\penalty0 e175, October 2013.
\newblock ISSN 2167-8359.
\newblock \doi{10.7717/peerj.175}.
\newblock URL \url{https://peerj.com/articles/175}.

\bibitem[Henneken and Accomazzi(2011)]{henneken2011linking}
Edwin~A Henneken and Alberto Accomazzi.
\newblock Linking to data-effect on citation rates in astronomy.
\newblock \emph{arXiv preprint arXiv:1111.3618}, 2011.

\bibitem[Sears(2011)]{sears2011data}
JR~Sears.
\newblock Data sharing effect on article citation rate in paleoceanography.
\newblock \emph{EOS, Transactions, American Geophysical Union}, 92\penalty0 (53):\penalty0 IN53B--1628, 2011.

\bibitem[Gleditsch et~al.(2003)Gleditsch, Metelits, and Strand]{gleditsch2003posting}
Nils~Petter Gleditsch, Claire Metelits, and Havard Strand.
\newblock Posting your data: Will you be scooped or will you be famous.
\newblock \emph{International Studies Perspectives}, 4\penalty0 (1):\penalty0 89--97, 2003.

\bibitem[Pienta et~al.(2011)Pienta, Alter, and Lyle]{pienta2011enduring}
Amy Pienta, George Alter, and Jared Lyle.
\newblock The enduring value of social science research.
\newblock In \emph{8 th International Conference on Preservation of Digital Objects}, page 215, 2011.

\bibitem[Peng(2011)]{peng_reproducible_2011}
Roger~D. Peng.
\newblock Reproducible research in computational science.
\newblock \emph{Science}, 334\penalty0 (6060):\penalty0 1226--1227, December 2011.
\newblock ISSN 0036-8075.
\newblock \doi{10.1126/science.1213847}.
\newblock URL \url{https://www.science.org/doi/10.1126/science.1213847}.

\bibitem[Zhao et~al.(2018)Zhao, Yan, and Li]{zhao_data_2018}
Mengnan Zhao, Erjia Yan, and Kai Li.
\newblock Data set mentions and citations: {A} content analysis of full-text publications.
\newblock \emph{Journal of the Association for Information Science and Technology}, 69\penalty0 (1):\penalty0 32--46, 2018.
\newblock ISSN 2330-1643.
\newblock \doi{10.1002/asi.23919}.
\newblock URL \url{https://onlinelibrary.wiley.com/doi/abs/10.1002/asi.23919}.

\bibitem[Koesten et~al.(2020)Koesten, Vougiouklis, Simperl, and Groth]{koesten_dataset_2020}
Laura Koesten, Pavlos Vougiouklis, Elena Simperl, and Paul Groth.
\newblock Dataset {Reuse}: {Toward} {Translating} {Principles} to {Practice}.
\newblock \emph{Patterns}, 1\penalty0 (8), November 2020.
\newblock ISSN 2666-3899.
\newblock \doi{10.1016/j.patter.2020.100136}.
\newblock URL \url{https://www.cell.com/patterns/abstract/S2666-3899(20)30184-7}.

\bibitem[Khan et~al.(2021)Khan, Thelwall, and Kousha]{khan_measuring_2021}
Nushrat Khan, Mike Thelwall, and Kayvan Kousha.
\newblock Measuring the impact of biodiversity datasets: data reuse, citations and altmetrics.
\newblock \emph{Scientometrics}, 126\penalty0 (4):\penalty0 3621--3639, February 2021.
\newblock ISSN 0138-9130.
\newblock \doi{10.1007/s11192-021-03890-6}.
\newblock URL \url{https://link.springer.com/10.1007/s11192-021-03890-6}.

\bibitem[Stodden et~al.(2018)Stodden, Seiler, and Ma]{stodden_empirical_2018}
Victoria Stodden, Jennifer Seiler, and Zhaokun Ma.
\newblock An empirical analysis of journal policy effectiveness for computational reproducibility.
\newblock \emph{Proceedings of the National Academy of Sciences}, 115\penalty0 (11):\penalty0 2584--2589, March 2018.
\newblock ISSN 0027-8424.
\newblock \doi{10.1073/pnas.1708290115}.
\newblock URL \url{https://pnas.org/doi/full/10.1073/pnas.1708290115}.

\bibitem[Cao et~al.(2023)Cao, Dodge, Lo, McFarland, and Wang]{cao_rise_2023}
Hancheng Cao, Jesse Dodge, Kyle Lo, Daniel~A. McFarland, and Lucy~Lu Wang.
\newblock The {Rise} of {Open} {Science}: {Tracking} the {Evolution} and {Perceived} {Value} of {Data} and {Methods} {Link}-{Sharing} {Practices}, October 2023.
\newblock URL \url{http://arxiv.org/abs/2310.03193}.

\bibitem[Colavizza et~al.(2020)Colavizza, Hrynaszkiewicz, Staden, Whitaker, and McGillivray]{colavizza_citation_2020}
Giovanni Colavizza, Iain Hrynaszkiewicz, Isla Staden, Kirstie Whitaker, and Barbara McGillivray.
\newblock The citation advantage of linking publications to research data.
\newblock \emph{PLOS ONE}, 15\penalty0 (4):\penalty0 e0230416, April 2020.
\newblock ISSN 1932-6203.
\newblock \doi{10.1371/journal.pone.0230416}.
\newblock URL \url{https://dx.plos.org/10.1371/journal.pone.0230416}.

\bibitem[Jiao et~al.(2024)Jiao, Li, and Fang]{jiao_data_2024}
Chenyue Jiao, Kai Li, and Zhichao Fang.
\newblock Data sharing practices across knowledge domains: {A} dynamic examination of data availability statements in {PLOS} {ONE} publications.
\newblock \emph{Journal of Information Science}, 50\penalty0 (3):\penalty0 673--689, June 2024.
\newblock ISSN 0165-5515.
\newblock \doi{10.1177/01655515221101830}.
\newblock URL \url{https://doi.org/10.1177/01655515221101830}.

\bibitem[Strcic et~al.(2022)Strcic, Civljak, Glozinic, Pacheco, Brkovic, and Puljak]{strcic_open_2022}
Josip Strcic, Antonia Civljak, Terezija Glozinic, Rafael~Leite Pacheco, Tonci Brkovic, and Livia Puljak.
\newblock Open data and data sharing in articles about {COVID}-19 published in preprint servers {medRxiv} and {bioRxiv}.
\newblock \emph{Scientometrics}, 127\penalty0 (5):\penalty0 2791--2802, March 2022.
\newblock ISSN 0138-9130.
\newblock \doi{10.1007/s11192-022-04346-1}.
\newblock URL \url{https://link.springer.com/10.1007/s11192-022-04346-1}.

\bibitem[Jahn et~al.(2023)Jahn, Laakso, Lazzeri, and McQuilton]{jahn_2023_7728016}
Najko Jahn, Mikael Laakso, Emma Lazzeri, and Peter McQuilton.
\newblock {Study on the readiness of research data and literature repositories to facilitate compliance with the Open Science Horizon Europe MGA requirements}, March 2023.
\newblock URL \url{https://doi.org/10.5281/zenodo.7728016}.

\bibitem[Commission et~al.(2021)Commission, for Research, and Innovation]{eu_openscience}
European Commission, Directorate-General for Research, and Innovation.
\newblock \emph{Horizon Europe, open science : early knowledge and data sharing, and open collaboration}.
\newblock Publications Office of the European Union, 2021.
\newblock \doi{doi/10.2777/18252}.

\bibitem[Robson et~al.(2021)Robson, Baum, Beaudry, Beitner, Brohmer, Chin, Jasko, Kouros, Laukkonen, Moreau, et~al.]{robson2021promoting}
Samuel~G Robson, Myriam~A Baum, Jennifer~L Beaudry, Julia Beitner, Hilmar Brohmer, Jason~M Chin, Katarzyna Jasko, Chrystyna~D Kouros, Ruben~E Laukkonen, David Moreau, et~al.
\newblock Promoting open science: a holistic approach to changing behaviour.
\newblock \emph{Collabra: Psychology}, 7\penalty0 (1):\penalty0 30137, 2021.

\bibitem[Gorgolewski and Poldrack(2016)]{gorgolewski2016practical}
Krzysztof~J Gorgolewski and Russell~A Poldrack.
\newblock A practical guide for improving transparency and reproducibility in neuroimaging research.
\newblock \emph{PLoS biology}, 14\penalty0 (7):\penalty0 e1002506, 2016.

\bibitem[Armeni et~al.(2021)Armeni, Brinkman, Carlsson, Eerland, Fijten, Fondberg, Heininga, Heunis, Koh, Masselink, et~al.]{armeni2021towards}
Kristijan Armeni, Loek Brinkman, Rickard Carlsson, Anita Eerland, Rianne Fijten, Robin Fondberg, Vera~E Heininga, Stephan Heunis, Wei~Qi Koh, Maurits Masselink, et~al.
\newblock Towards wide-scale adoption of open science practices: The role of open science communities.
\newblock \emph{Science and Public Policy}, 48\penalty0 (5):\penalty0 605--611, 2021.

\bibitem[Gopal et~al.(2018)Gopal, Wallach, Aminawung, Gonsalves, Dal-R{\'e}, Miller, and Ross]{gopal2018adherence}
Anand~D Gopal, Joshua~D Wallach, Jenerius~A Aminawung, Gregg Gonsalves, Rafael Dal-R{\'e}, Jennifer~E Miller, and Joseph~S Ross.
\newblock Adherence to the international committee of medical journal editors’(icmje) prospective registration policy and implications for outcome integrity: a cross-sectional analysis of trials published in high-impact specialty society journals.
\newblock \emph{Trials}, 19:\penalty0 1--13, 2018.

\bibitem[McKiernan et~al.(2016)McKiernan, Bourne, Brown, Buck, Kenall, Lin, McDougall, Nosek, Ram, Soderberg, Spies, Thaney, Updegrove, Woo, and Yarkoni]{mckiernan_how_2016}
Erin~C McKiernan, Philip~E Bourne, C~Titus Brown, Stuart Buck, Amye Kenall, Jennifer Lin, Damon McDougall, Brian~A Nosek, Karthik Ram, Courtney~K Soderberg, Jeffrey~R Spies, Kaitlin Thaney, Andrew Updegrove, Kara~H Woo, and Tal Yarkoni.
\newblock How open science helps researchers succeed.
\newblock \emph{eLife}, 5:\penalty0 e16800, July 2016.
\newblock ISSN 2050-084X.
\newblock \doi{10.7554/eLife.16800}.
\newblock URL \url{https://elifesciences.org/articles/16800}.

\bibitem[Huang et~al.(2024)Huang, Neylon, Montgomery, Hosking, Diprose, Handcock, and Wilson]{huang2024open}
Chun-Kai Huang, Cameron Neylon, Lucy Montgomery, Richard Hosking, James~P Diprose, Rebecca~N Handcock, and Katie Wilson.
\newblock Open access research outputs receive more diverse citations.
\newblock \emph{Scientometrics}, 129\penalty0 (2):\penalty0 825--845, 2024.

\bibitem[Schultz(2021)]{schultz2021all}
Teresa Schultz.
\newblock All the research that’s fit to print: Open access and the news media.
\newblock \emph{Quantitative Science Studies}, 2\penalty0 (3):\penalty0 828--844, 2021.

\bibitem[Yang et~al.(2024)Yang, Shoaib, West, and Colavizza]{yang2024open}
Puyu Yang, Ahad Shoaib, Robert West, and Giovanni Colavizza.
\newblock Open access improves the dissemination of science: insights from wikipedia.
\newblock \emph{Scientometrics}, pages 1--24, 2024.

\bibitem[Woelfle et~al.(2011)Woelfle, Olliaro, and Todd]{woelfle2011open}
Michael Woelfle, Piero Olliaro, and Matthew~H Todd.
\newblock Open science is a research accelerator.
\newblock \emph{Nature chemistry}, 3\penalty0 (10):\penalty0 745--748, 2011.

\bibitem[Besan{\c{c}}on et~al.(2021)Besan{\c{c}}on, Peiffer-Smadja, Segalas, Jiang, Masuzzo, Smout, Billy, Deforet, and Leyrat]{besanccon2021open}
Lonni Besan{\c{c}}on, Nathan Peiffer-Smadja, Corentin Segalas, Haiting Jiang, Paola Masuzzo, Cooper Smout, Eric Billy, Maxime Deforet, and Cl{\'e}mence Leyrat.
\newblock Open science saves lives: lessons from the covid-19 pandemic.
\newblock \emph{BMC Medical Research Methodology}, 21\penalty0 (1):\penalty0 117, 2021.

\bibitem[Collaboration(2015)]{open2015estimating}
Open~Science Collaboration.
\newblock Estimating the reproducibility of psychological science.
\newblock \emph{Science}, 349\penalty0 (6251):\penalty0 aac4716, 2015.

\bibitem[Hossain et~al.(2016)Hossain, Dwivedi, and Rana]{hossain2016state}
Mohammad~Alamgir Hossain, Yogesh~K Dwivedi, and Nripendra~P Rana.
\newblock State-of-the-art in open data research: Insights from existing literature and a research agenda.
\newblock \emph{Journal of organizational computing and electronic commerce}, 26\penalty0 (1-2):\penalty0 14--40, 2016.

\bibitem[Milham et~al.(2018)Milham, Craddock, Son, Fleischmann, Clucas, Xu, Koo, Krishnakumar, Biswal, Castellanos, Colcombe, Di~Martino, Zuo, and Klein]{milham_assessment_2018}
Michael~P. Milham, R.~Cameron Craddock, Jake~J. Son, Michael Fleischmann, Jon Clucas, Helen Xu, Bonhwang Koo, Anirudh Krishnakumar, Bharat~B. Biswal, F.~Xavier Castellanos, Stan Colcombe, Adriana Di~Martino, Xi-Nian Zuo, and Arno Klein.
\newblock Assessment of the impact of shared brain imaging data on the scientific literature.
\newblock \emph{Nature Communications}, 9\penalty0 (1):\penalty0 2818, July 2018.
\newblock ISSN 2041-1723.
\newblock \doi{10.1038/s41467-018-04976-1}.
\newblock URL \url{https://www.nature.com/articles/s41467-018-04976-1}.

\bibitem[Shahin et~al.(2020)Shahin, Bhattacharya, Silva, Kim, Burton, Podichetty, Romero, and Conrado]{shahin2020open}
Mohamed~H Shahin, Sanchita Bhattacharya, Diego Silva, Sarah Kim, Jackson Burton, Jagdeep Podichetty, Klaus Romero, and Daniela~J Conrado.
\newblock Open data revolution in clinical research: opportunities and challenges.
\newblock \emph{Clinical and Translational Science}, 13\penalty0 (4):\penalty0 665--674, 2020.

\bibitem[Los(2010)]{los_riding_2010}
Wouter Los.
\newblock \emph{Riding the wave {How} {Europe} can gain from the rising tide of scientific data {Final} report of the {High} {Level} {Expert} {Group} on {Scientific} {Data} {A} submission to the {European} {Commission}}.
\newblock European Union, January 2010.

\bibitem[Bromley(1991)]{bromley_policy_1991}
Allan Bromley.
\newblock Policy {Statements} on {Data} {Management} for {Global} {Change} {Research}, February 1991.
\newblock URL \url{https://digital.library.unt.edu/ark:/67531/metadc11862/}.

\bibitem[{General Office of the State Council of the People's Republic of China}(2018)]{china____policy}
{General Office of the State Council of the People's Republic of China}.
\newblock Notification by the general office of the state council on the issuance of scientific data management practices., 2018.
\newblock URL \url{https://www.gov.cn/zhengce/content/2018-04/02/content_5279272.htm}.

\bibitem[UK(2016)]{uk_policy}
Research~Councils UK.
\newblock Concordat on open research data, 2016.
\newblock URL \url{https://www.ukri.org/wp-content/uploads/2020/10/UKRI-020920-ConcordatonOpenResearchData.pdf}.

\bibitem[Service(2011)]{australia_policy}
Australian National~Data Service.
\newblock Outline of a research data management policy for australian universities / institutions, 2011.
\newblock URL \url{https://alliancecan.ca/sites/default/files/2022-03/institutional-research-data-management-policies.pdf}.

\bibitem[Smith(2012)]{smith2012institutional}
Mackenzie Smith.
\newblock Institutional perspectives on credit systems for research data.
\newblock In \emph{For attribution: Developing scientific data attribution and citation practices and standards: Summary of an international workshop}, pages 77--80, 2012.

\bibitem[Spengler(2012)]{spengler2012data}
S~Spengler.
\newblock Data citation and attribution: A funder’s perspective.
\newblock In \emph{For attribution: Developing scientific data attribution and citation practices and standards: Summary of an international workshop}, pages 177--188, 2012.

\bibitem[Cousijn et~al.(2018)Cousijn, Kenall, Ganley, Harrison, Kernohan, Lemberger, Murphy, Polischuk, Taylor, Martone, and Clark]{cousijn_data_2018}
Helena Cousijn, Amye Kenall, Emma Ganley, Melissa Harrison, David Kernohan, Thomas Lemberger, Fiona Murphy, Patrick Polischuk, Simone Taylor, Maryann Martone, and Tim Clark.
\newblock A data citation roadmap for scientific publishers.
\newblock \emph{Scientific Data}, 5\penalty0 (1):\penalty0 180259, November 2018.
\newblock ISSN 2052-4463.
\newblock \doi{10.1038/sdata.2018.259}.
\newblock URL \url{https://www.nature.com/articles/sdata2018259}.

\bibitem[Walton(2010)]{walton_data_2010}
David W.~H. Walton.
\newblock Data {Citation} - {Moving} to {New} {Norms}.
\newblock \emph{Antarctic Science}, 22\penalty0 (4):\penalty0 333--333, August 2010.
\newblock ISSN 0954-1020.
\newblock \doi{10.1017/S0954102010000520}.
\newblock URL \url{https://www.cambridge.org/core/product/identifier/S0954102010000520/type/journal_article}.

\bibitem[{PLOS One}(2019)]{plos_policy}
{PLOS One}.
\newblock Plos one data availability, December 2019.
\newblock URL \url{https://journals.plos.org/plosone/s/data-availability}.

\bibitem[{Springer Nature}(2016)]{Springer_policy}
{Springer Nature}.
\newblock Research data policy, 2016.
\newblock URL \url{https://www.springernature.com/gp/authors/research-data-policy}.

\bibitem[Niemeyer et~al.(2016)Niemeyer, Smith, and Katz]{niemeyer_challenge_2016}
Kyle~E. Niemeyer, Arfon~M. Smith, and Daniel~S. Katz.
\newblock The {Challenge} and {Promise} of {Software} {Citation} for {Credit}, {Identification}, {Discovery}, and {Reuse}.
\newblock \emph{Journal of Data and Information Quality}, 7\penalty0 (4):\penalty0 1--5, October 2016.
\newblock ISSN 1936-1955.
\newblock \doi{10.1145/2968452}.
\newblock URL \url{https://dl.acm.org/doi/10.1145/2968452}.

\bibitem[McNaught(2015)]{mcnaught_changing_2015}
Keith McNaught.
\newblock The {Changing} {Publication} {Practices} in {Academia}: {Inherent} {Uses} and {Issues} in {Open} {Access} and {Online} {Publishing} and the {Rise} of {Fraudulent} {Publications}.
\newblock \emph{The Journal of Electronic Publishing}, 18\penalty0 (3), June 2015.
\newblock ISSN 1080-2711.
\newblock \doi{10.3998/3336451.0018.308}.
\newblock URL \url{http://hdl.handle.net/2027/spo.3336451.0018.308}.

\bibitem[Belter(2014)]{belter_measuring_2014}
Christopher~W. Belter.
\newblock Measuring the {Value} of {Research} {Data}: {A} {Citation} {Analysis} of {Oceanographic} {Data} {Sets}.
\newblock \emph{PLoS ONE}, 9\penalty0 (3):\penalty0 e92590, March 2014.
\newblock ISSN 1932-6203.
\newblock \doi{10.1371/journal.pone.0092590}.
\newblock URL \url{https://dx.plos.org/10.1371/journal.pone.0092590}.

\bibitem[Chawinga and Zinn(2019)]{chawinga_global_2019}
Winner~Dominic Chawinga and Sandy Zinn.
\newblock Global perspectives of research data sharing: {A} systematic literature review.
\newblock \emph{Library \& Information Science Research}, 41\penalty0 (2):\penalty0 109--122, April 2019.
\newblock ISSN 0740-8188.
\newblock \doi{10.1016/j.lisr.2019.04.004}.
\newblock URL \url{https://www.sciencedirect.com/science/article/pii/S074081881830330X}.

\bibitem[Gajbe et~al.(2021)Gajbe, Tiwari, {Gopalji}, and Singh]{gajbe_evaluation_2021}
Sagar~Bhimrao Gajbe, Amit Tiwari, {Gopalji}, and Ranjeet~Kumar Singh.
\newblock Evaluation and analysis of {Data} {Management} {Plan} tools: {A} parametric approach.
\newblock \emph{Information Processing \& Management}, 58\penalty0 (3):\penalty0 102480, May 2021.
\newblock ISSN 0306-4573.
\newblock \doi{10.1016/j.ipm.2020.102480}.
\newblock URL \url{https://linkinghub.elsevier.com/retrieve/pii/S0306457320309699}.

\bibitem[Tenopir et~al.(2011)Tenopir, Allard, Douglass, Aydinoglu, Wu, Read, Manoff, and Frame]{tenopir_data_2011}
Carol Tenopir, Suzie Allard, Kimberly Douglass, Arsev~Umur Aydinoglu, Lei Wu, Eleanor Read, Maribeth Manoff, and Mike Frame.
\newblock Data {Sharing} by {Scientists}: {Practices} and {Perceptions}.
\newblock \emph{PLOS ONE}, 6\penalty0 (6):\penalty0 e21101, June 2011.
\newblock ISSN 1932-6203.
\newblock \doi{10.1371/journal.pone.0021101}.
\newblock URL \url{https://dx.plos.org/10.1371/journal.pone.0021101}.

\bibitem[Tenopir et~al.(2020)Tenopir, Rice, Allard, Baird, Borycz, Christian, Grant, Olendorf, and Sandusky]{tenopir_data_2020}
Carol Tenopir, Natalie~M. Rice, Suzie Allard, Lynn Baird, Josh Borycz, Lisa Christian, Bruce Grant, Robert Olendorf, and Robert~J. Sandusky.
\newblock Data sharing, management, use, and reuse: {Practices} and perceptions of scientists worldwide.
\newblock \emph{PLOS ONE}, 15\penalty0 (3):\penalty0 e0229003, March 2020.
\newblock ISSN 1932-6203.
\newblock \doi{10.1371/journal.pone.0229003}.
\newblock URL \url{https://dx.plos.org/10.1371/journal.pone.0229003}.

\bibitem[Markiewicz et~al.(2021)Markiewicz, Gorgolewski, Feingold, Blair, Halchenko, Miller, Hardcastle, Wexler, Esteban, Goncavles, Jwa, and Poldrack]{markiewicz_openneuro_2021}
Christopher~J Markiewicz, Krzysztof~J Gorgolewski, Franklin Feingold, Ross Blair, Yaroslav~O Halchenko, Eric Miller, Nell Hardcastle, Joe Wexler, Oscar Esteban, Mathias Goncavles, Anita Jwa, and Russell Poldrack.
\newblock The {OpenNeuro} resource for sharing of neuroscience data.
\newblock \emph{eLife}, 10:\penalty0 e71774, October 2021.
\newblock ISSN 2050-084X.
\newblock \doi{10.7554/eLife.71774}.
\newblock URL \url{https://elifesciences.org/articles/71774}.

\bibitem[Tedersoo et~al.(2021)Tedersoo, Küngas, Oras, Köster, Eenmaa, Leijen, Pedaste, Raju, Astapova, Lukner, Kogermann, and Sepp]{tedersoo_data_2021}
Leho Tedersoo, Rainer Küngas, Ester Oras, Kajar Köster, Helen Eenmaa, Äli Leijen, Margus Pedaste, Marju Raju, Anastasiya Astapova, Heli Lukner, Karin Kogermann, and Tuul Sepp.
\newblock Data sharing practices and data availability upon request differ across scientific disciplines.
\newblock \emph{Scientific Data}, 8\penalty0 (1):\penalty0 192, July 2021.
\newblock ISSN 2052-4463.
\newblock \doi{10.1038/s41597-021-00981-0}.
\newblock URL \url{https://www.nature.com/articles/s41597-021-00981-0}.

\bibitem[Danchev et~al.(2021)Danchev, Min, Borghi, Baiocchi, and Ioannidis]{danchev_evaluation_2021}
Valentin Danchev, Yan Min, John Borghi, Mike Baiocchi, and John P.~A. Ioannidis.
\newblock Evaluation of {Data} {Sharing} {After} {Implementation} of the {International} {Committee} of {Medical} {Journal} {Editors} {Data} {Sharing} {Statement} {Requirement}.
\newblock \emph{JAMA Network Open}, 4\penalty0 (1):\penalty0 e2033972, January 2021.
\newblock ISSN 2574-3805.
\newblock \doi{10.1001/jamanetworkopen.2020.33972}.
\newblock URL \url{https://jamanetwork.com/journals/jamanetworkopen/fullarticle/2775667}.

\bibitem[Park et~al.(2018)Park, You, and Wolfram]{park_informal_2018}
Hyoungjoo Park, Sukjin You, and Dietmar Wolfram.
\newblock Informal data citation for data sharing and reuse is more common than formal data citation in biomedical fields.
\newblock \emph{Journal of the Association for Information Science and Technology}, 69\penalty0 (11):\penalty0 1346--1354, 2018.
\newblock ISSN 2330-1643.
\newblock \doi{10.1002/asi.24049}.
\newblock URL \url{https://onlinelibrary.wiley.com/doi/abs/10.1002/asi.24049}.

\bibitem[Robinson-García et~al.(2016)Robinson-García, Jiménez-Contreras, and Torres-Salinas]{robinson-garcia_analyzing_2016}
Nicolas Robinson-García, Evaristo Jiménez-Contreras, and Daniel Torres-Salinas.
\newblock Analyzing data citation practices using the data citation index.
\newblock \emph{Journal of the Association for Information Science and Technology}, 67\penalty0 (12):\penalty0 2964--2975, 2016.
\newblock ISSN 2330-1643.
\newblock \doi{10.1002/asi.23529}.
\newblock URL \url{https://onlinelibrary.wiley.com/doi/abs/10.1002/asi.23529}.

\bibitem[Park and Wolfram(2017)]{park_examination_2017}
Hyoungjoo Park and Dietmar Wolfram.
\newblock An examination of research data sharing and re-use: implications for data citation practice.
\newblock \emph{Scientometrics}, 111\penalty0 (1):\penalty0 443--461, April 2017.
\newblock ISSN 1588-2861.
\newblock \doi{10.1007/s11192-017-2240-2}.
\newblock URL \url{https://doi.org/10.1007/s11192-017-2240-2}.

\bibitem[Piwowar et~al.(2007)Piwowar, Day, and Fridsma]{piwowar_sharing_2007}
Heather~A. Piwowar, Roger~S. Day, and Douglas~B. Fridsma.
\newblock Sharing {Detailed} {Research} {Data} {Is} {Associated} with {Increased} {Citation} {Rate}.
\newblock \emph{PLOS ONE}, 2\penalty0 (3):\penalty0 e308, March 2007.
\newblock ISSN 1932-6203.
\newblock \doi{10.1371/journal.pone.0000308}.
\newblock URL \url{https://dx.plos.org/10.1371/journal.pone.0000308}.

\bibitem[Helbig et~al.(2015)Helbig, Hausstein, and Toepfer]{helbig_supporting_2015}
Kerstin Helbig, Brigitte Hausstein, and Ralf Toepfer.
\newblock Supporting {Data} {Citation}: {Experiences} and {Best} {Practices} of a {DOI} {Allocation} {Agency} for {Social} {Sciences}.
\newblock \emph{Journal of Librarianship and Scholarly Communication}, 3\penalty0 (2):\penalty0 1220, September 2015.
\newblock ISSN 2162-3309.
\newblock \doi{10.7710/2162-3309.1220}.
\newblock URL \url{https://jlsc-pub.org/article/10.7710/2162-3309.1220/}.

\bibitem[Torres-Salinas et~al.(2014)Torres-Salinas, Jiménez-Contreras, and Robinson-García]{torres-salinas_how_2014}
Daniel Torres-Salinas, Evaristo Jiménez-Contreras, and Nicolas Robinson-García.
\newblock How many citations are there in the {Data} {Citation} {Index}?, September 2014.
\newblock URL \url{http://arxiv.org/abs/1409.0753}.

\bibitem[Tenopir et~al.(2015)Tenopir, Dalton, Allard, Frame, Pjesivac, Birch, Pollock, and Dorsett]{tenopir_changes_2015}
Carol Tenopir, Elizabeth~D. Dalton, Suzie Allard, Mike Frame, Ivanka Pjesivac, Ben Birch, Danielle Pollock, and Kristina Dorsett.
\newblock Changes in {Data} {Sharing} and {Data} {Reuse} {Practices} and {Perceptions} among {Scientists} {Worldwide}.
\newblock \emph{PLOS ONE}, 10\penalty0 (8):\penalty0 e0134826, August 2015.
\newblock ISSN 1932-6203.
\newblock \doi{10.1371/journal.pone.0134826}.
\newblock URL \url{https://dx.plos.org/10.1371/journal.pone.0134826}.

\bibitem[Peters et~al.(2015)Peters, Kraker, Lex, Gumpenberger, and Gorraiz]{peters_research_2015}
Isabella Peters, Peter Kraker, Elisabeth Lex, Christian Gumpenberger, and Juan Gorraiz.
\newblock Research {Data} {Explored}: {Citations} versus {Altmetrics}, April 2015.
\newblock URL \url{http://arxiv.org/abs/1501.03342}.

\bibitem[Kafkas et~al.(2015)Kafkas, Kim, Pi, and McEntyre]{kafkas_database_2015}
Senay Kafkas, Jee-Hyub Kim, Xingjun Pi, and Johanna~R. McEntyre.
\newblock Database citation in supplementary data linked to {Europe} {PubMed} {Central} full text biomedical articles.
\newblock \emph{Journal of Biomedical Semantics}, 6\penalty0 (1):\penalty0 1, 2015.
\newblock ISSN 2041-1480.
\newblock \doi{10.1186/2041-1480-6-1}.
\newblock URL \url{http://www.jbiomedsem.com/content/6/1/1}.

\bibitem[Piwowar et~al.(2011)Piwowar, Carlson, and Vision]{piwowar_beginning_2011}
Heather~A. Piwowar, Jonathan~D. Carlson, and Todd~J. Vision.
\newblock Beginning to track 1000 datasets from public repositories into the published literature.
\newblock \emph{Proceedings of the American Society for Information Science and Technology}, 48\penalty0 (1):\penalty0 1--4, 2011.
\newblock ISSN 0044-7870.
\newblock \doi{10.1002/meet.2011.14504801337}.
\newblock URL \url{https://onlinelibrary.wiley.com/doi/10.1002/meet.2011.14504801337}.

\bibitem[Silvello(2017)]{silvello_theory_2017}
Gianmaria Silvello.
\newblock Theory and practice of data citation.
\newblock \emph{Journal of the Association for Information Science and Technology}, 69\penalty0 (1):\penalty0 6--20, September 2017.
\newblock ISSN 2330-1635.
\newblock \doi{10.1002/asi.23917}.
\newblock URL \url{https://asistdl.onlinelibrary.wiley.com/doi/10.1002/asi.23917}.

\bibitem[Federer(2022)]{federer_long-term_2022}
Lisa~M. Federer.
\newblock Long-term availability of data associated with articles in {PLOS} {ONE}.
\newblock \emph{PLOS ONE}, 17\penalty0 (8):\penalty0 e0272845, August 2022.
\newblock ISSN 1932-6203.
\newblock \doi{10.1371/journal.pone.0272845}.
\newblock URL \url{https://dx.plos.org/10.1371/journal.pone.0272845}.

\bibitem[Zhang et~al.(2022)Zhang, Zhao, Wang, Chen, Mahmood, Zaib, Zhang, and Sheng]{zhang2022towards}
Yang Zhang, Rongying Zhao, Yufei Wang, Haihua Chen, Adnan Mahmood, Munazza Zaib, Wei~Emma Zhang, and Quan~Z Sheng.
\newblock Towards employing native information in citation function classification.
\newblock \emph{Scientometrics}, pages 1--21, 2022.

\bibitem[Devlin(2018)]{devlin2018bert}
Jacob Devlin.
\newblock Bert: Pre-training of deep bidirectional transformers for language understanding.
\newblock \emph{arXiv preprint arXiv:1810.04805}, 2018.
\newblock URL \url{https://arxiv.org/abs/1810.04805}.

\bibitem[Van~Eck and Waltman(2010)]{van2010software}
Nees Van~Eck and Ludo Waltman.
\newblock Software survey: Vosviewer, a computer program for bibliometric mapping.
\newblock \emph{scientometrics}, 84\penalty0 (2):\penalty0 523--538, 2010.

\bibitem[Jim{\'e}nez-Contreras et~al.(2015)Jim{\'e}nez-Contreras, Torres-Salinas, and Robinson-Garcia]{jimenez2015analyzing}
Evaristo Jim{\'e}nez-Contreras, Daniel Torres-Salinas, and Nicolas Robinson-Garcia.
\newblock Analyzing data citation practices using the data citation index.
\newblock \emph{Journal of the Association for Information Science and Technology}, 2015.

\bibitem[Huang et~al.(2012)Huang, Hawkins, Lei, Miller, Favret, Zhang, and Qiao]{huang2012willing}
Xiaolei Huang, Bradford~A Hawkins, Fumin Lei, Gary~L Miller, Colin Favret, Ruiling Zhang, and Gexia Qiao.
\newblock Willing or unwilling to share primary biodiversity data: results and implications of an international survey.
\newblock \emph{Conservation letters}, 5\penalty0 (5):\penalty0 399--406, 2012.

\bibitem[Oushy et~al.(2015)Oushy, Palacios, Holden, Ramirez, Gallion, and O’Connell]{oushy2015share}
Mai~H Oushy, Rebecca Palacios, Alan~EC Holden, Amelie~G Ramirez, Kipling~J Gallion, and Mary~A O’Connell.
\newblock To share or not to share? a survey of biomedical researchers in the us southwest, an ethnically diverse region.
\newblock \emph{PLoS One}, 10\penalty0 (9):\penalty0 e0138239, 2015.

\bibitem[Kim and Yoon(2017)]{kim2017scientists}
Youngseek Kim and Ayoung Yoon.
\newblock Scientists' data reuse behaviors: A multilevel analysis.
\newblock \emph{Journal of the Association for Information Science and Technology}, 68\penalty0 (12):\penalty0 2709--2719, 2017.

\bibitem[Force(2014)]{force2014force}
II~Force.
\newblock Force ii: The future of research communication and scholarship, joint declaration of data citation principles, 2014.
\newblock URL \url{https://force11.org/info/joint-declaration-of-data-citation-principles-final/}.

\bibitem[Kratz and Strasser(2014)]{kratz2014data}
John Kratz and Carly Strasser.
\newblock Data publication consensus and controversies.
\newblock \emph{F1000Research}, 3, 2014.

\bibitem[Altman et~al.(2015)Altman, Borgman, Crosas, and Matone]{altman2015introduction}
Micah Altman, Christine Borgman, Merc{\`e} Crosas, and Maryann Matone.
\newblock An introduction to the joint principles for data citation.
\newblock \emph{Bulletin of the Association for Information Science and Technology}, 41\penalty0 (3):\penalty0 43--45, 2015.

\bibitem[Park(2022)]{park2022interdisciplinarity}
Hyoungjoo Park.
\newblock The interdisciplinarity of research data: How widely is shared research data reused in the stem fields?
\newblock \emph{The Journal of Academic Librarianship}, 48\penalty0 (4):\penalty0 102535, 2022.

\bibitem[Holdren et~al.(2013)]{holdren2013memorandum}
John~P Holdren et~al.
\newblock Memorandum for the heads of executive departments and agencies: Increasing access to the results of federally funded scientific research, 2013.
\newblock URL \url{https://rosap.ntl.bts.gov/view/dot/34953/dot_34953_DS1.pdf}.

\bibitem[Schiltz(2018)]{plans2018}
Marc Schiltz.
\newblock Plan s, 2018.
\newblock URL \url{https://www.coalition-s.org/plan-s-funders-implementation/}.

\bibitem[F{\"u}hr and Bisset~Alvarez(2021)]{fuhr2021digital}
Fabiane F{\"u}hr and Edgar Bisset~Alvarez.
\newblock Digital humanities and open science: initial aspects.
\newblock In \emph{International Conference on Data and Information in Online}, pages 154--173. Springer, 2021.

\bibitem[Borgman(2012)]{borgman_conundrum_2012}
Christine~L. Borgman.
\newblock The conundrum of sharing research data.
\newblock \emph{Journal of the American Society for Information Science and Technology}, 63\penalty0 (6):\penalty0 1059--1078, April 2012.
\newblock ISSN 1532-2882.
\newblock \doi{10.1002/asi.22634}.
\newblock URL \url{https://onlinelibrary.wiley.com/doi/10.1002/asi.22634}.

\bibitem[Mayernik(2017)]{mayernik2017open}
Matthew~S Mayernik.
\newblock Open data: Accountability and transparency.
\newblock \emph{Big Data \& Society}, 4\penalty0 (2):\penalty0 2053951717718853, 2017.

\end{thebibliography}






\end{document}